\documentclass[12pt]{iopart}

\expandafter\let\csname equation*\endcsname=\relax
\expandafter\let\csname endequation*\endcsname=\relax

\usepackage{amsmath}
\usepackage{enumerate}
\usepackage{graphicx}
\usepackage{mathbbol}
\usepackage{amsfonts}

\newcommand{\lmax}{\lambda_{\max}}

\def\Xint#1{\mathchoice
   {\XXint\displaystyle\textstyle{#1}}%
   {\XXint\textstyle\scriptstyle{#1}}%
   {\XXint\scriptstyle\scriptscriptstyle{#1}}%
   {\XXint\scriptscriptstyle\scriptscriptstyle{#1}}%
   \!\int}
\def\XXint#1#2#3{{\setbox0=\hbox{$#1{#2#3}{\int}$}
     \vcenter{\hbox{$#2#3$}}\kern-.5\wd0}}

\def\dashint{\Xint-}

\begin{document}
\title[Top eigenvalue of a random matrix: large deviations and $3^{\rm rd}$ order transition]{Top eigenvalue of a random matrix: large deviations and third order phase transition}

\author{Satya N. Majumdar}
\address{Universit\'e Paris-Sud, LPTMS, CNRS (UMR 8626), 91405 Orsay Cedex, France}

\author{Gr\'egory Schehr}
\address{Universit\'e Paris-Sud, LPTMS, CNRS (UMR 8626), 91405 Orsay Cedex, France}

\begin{abstract}
We study the fluctuations of the largest eigenvalue $\lmax$ of $N \times N$ random matrices in the limit of large $N$. The main focus is on Gaussian $\beta$-ensembles, including in particular the Gaussian orthogonal ($\beta=1$), unitary ($\beta=2$) and symplectic ($\beta = 4$) ensembles. The probability density function (PDF) of $\lmax$ consists, for large $N$, of a central part described by Tracy-Widom distributions flanked, on both sides, by two large deviations tails. While the central part characterizes the typical fluctuations of $\lmax$ -- of order ${\cal O}(N^{-2/3})$ --, the large deviations tails are instead associated to extremely rare fluctuations -- of order ${\cal O}(1)$. Here we review some recent developments in the theory of these extremely rare events using a Coulomb gas approach. We discuss in particular the third-order phase transition which separates the left tail from the right tail, a transition akin to the so-called Gross-Witten-Wadia phase transition found in $2$-d lattice quantum chromodynamics. We also discuss the occurrence of similar third-order transitions in various physical problems, including non-intersecting Brownian motions, conductance fluctuations in mesoscopic physics and entanglement in a bipartite system. 
\end{abstract}

\maketitle

\section{Introduction and motivations}

Since the pioneering work of Wishart in statistics~\cite{Wis28},
followed by Wigner and others in nuclear physics~\cite{Wig51,Meh91}, Random 
Matrix 
Theory (RMT) has found a huge number of applications ranging from 
statistical physics of disordered systems, mesoscopic physics, quantum 
information, finance, telecommunication networks to number theory, 
combinatorics, integrable systems and quantum chromodynamics (QCD), to name 
just a few~\cite{ABdF}. Among the recent developments in RMT, the study of 
the largest eigenvalue $\lmax$ of large random matrices has attracted  
particular attention. Questions related to the fluctuations of $\lmax$ 
belong to the wider topic of extreme value statistics (EVS). Being at the 
heart of optimization problems, such extreme value questions arise 
naturally in the statistical physics of complex and disordered systems 
\cite{BM97,DM01,LDM03}. In particular, the eigenvalues of a random matrix 
provide an interesting laboratory to study EVS of \emph{strongly 
correlated}
random variables, and go beyond the three standard universality classes of 
EVS for independent and identically distributed (i. i. d.) random 
variables~\cite{Gum58}.

As realized a long time ago in a seminal paper by May~\cite{May72}, 
a natural application of the statistics of $\lmax$ is to provide 
a criterion of 
physical stability in dynamical systems such as ecosystems. Near a 
fixed point 
of a dynamical system, one can linearize the equations of motion and the 
eigenvalues of the corresponding matrix associated with the linear 
equations provide important informations about the stability of the fixed 
point. For example, if all the eigenvalues are negative (or positive) the 
fixed point is stable (or unstable). As a concrete example, May
considered~\cite{May72} a population of $N$ 
distinct 
species with equilibrium densities $\rho_i^*$, $i= 
1, 2, 
\cdots, N$. To start with, they are \emph{noninteracting} and \emph{stable}
in the sense that when slightly perturbed from their equilibrium 
densities, each density relaxes to its equilibrium value with some 
characteristic 
damping time. For simplicity, these damping times are all chosen to 
be unity which sets the time scale. 
Hence the equations of motion for $x_i(t) = \rho_i(t) - \rho_i^*$, to 
linear order, are simply $d x_i(t)/dt = - x_i(t)$. Now, imagine
switching on pair-wise interactions between the species. 
May assumed that the interactions between pairs of species can be modeled
by a random matrix ${\bf J}$, of size $N \times N$, which is 
real and symmetric ($J_{ij} = J_{ji}$). The linearized equation of motions 
close to $\rho_i^*$ then read, in presence of interactions~\cite{May72}:
\begin{eqnarray}\label{linear_stability}
\frac{d x_i(t)}{dt} = - x_i(t) + \alpha \sum_{j=1}^N J_{ij} x_j(t) \;,
\end{eqnarray}
where $\alpha$ sets the strength of the interactions. A natural question 
is then: what is the probability, $P_{\rm stable}(\alpha,N)$, that the 
system described by (\ref{linear_stability}) remains stable once the 
interactions are switched on \cite{May72}? In other words,
what is the probability that the fixed point $x_i=0$ for all $i$ remains
stable in presence of a nonzero $\alpha$?
By transforming (\ref{linear_stability}) to the diagonal 
basis of
${\bf
J}$, it is easy to see that the fixed point $x_i=0$ will remain stable,
provided the eigenvalues $\lambda_i$ of  
the random matrix ${\bf 
J}$ satisfy the inequality: $\alpha \, \lambda_i - 1 \leq 0$, for all 
$i=1, \cdots, N$. This is equivalent to
the statement that the largest eigenvalue $\lmax = \max_{1\leq i \leq 
N}
\lambda_i$ satisfies the inequality: $\lmax\le 1/\alpha$. Hence the 
probability that the 
system in 
(\ref{linear_stability}) is stable gets naturally related to the 
cumulative 
distribution function (CDF) of the largest eigenvalue $\lmax$:
\begin{eqnarray}\label{rel_lmax}
P_{\rm stable}(\alpha,N) =  {\rm Prob.} \, [\lmax \leq w = 1/\alpha] \;.
\end{eqnarray}

\begin{figure}
\begin{center}
\includegraphics[width = 0.7\linewidth]{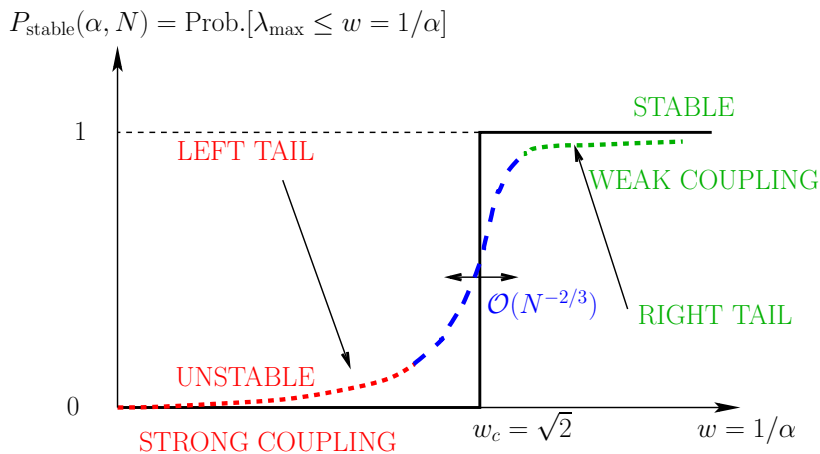}
\caption{Sketch of the diagram of stability of 
the system described by (\ref{linear_stability}). 
The solid line is the limit $N \to \infty$, illustrating the 
transition found by May~\cite{May72} at $w=w_c=\sqrt{2}$. The dashed line 
corresponds 
to the large but finite $N$ regime: the central regime 
(of order ${\cal O}(N^{-2/3})$  and described by 
Tracy-Widom distributions for Gaussian random matrices) is flanked 
by large deviations tails on both sides. Precise description of the left and 
the right tail is 
the main subject of the present article.}\label{fig:stability}
\end{center}
\end{figure}

Assuming that $J_{ij}$'s are identical Gaussian random 
variables with variance $1/N$, i.e. the matrix ${\mathbf J}$ belongs to 
the Gaussian orthogonal ensemble (GOE) of RMT, May 
noticed~\cite{May72} that this cumulative probability 
(\ref{rel_lmax}) undergoes a sharp transition (when $N \to \infty$) as 
$\alpha$ increases beyond the critical value $\alpha_c = 1/\sqrt{2}$ (see 
Fig. 
\ref{fig:stability}) [see also~\cite{GA70} 
for anterior numerical simulations]
\begin{eqnarray}\label{stable_transition}
\hspace*{-2cm}\lim_{N \to \infty} P_{\rm stable}(\alpha,N) = 
\begin{cases}
&1 \;, \; \alpha < \alpha_c \; : \;{\rm stable \;,\; weakly \; 
interacting \; phase}\\
&0 \;, \;  \alpha > \alpha_c \; : \; {\rm unstable \;, 
\; strongly \; interacting \; phase} \;.
\end{cases}
\end{eqnarray}

May's work was indeed the first direct physical application
of the statistics of $\lmax$ and to our knowledge, the first
one to point out the 
existence of a \emph{sharp} phase transition associated with the CDF of $\lmax$. Several questions
follow quite naturally. May's transition occurs strictly
in the $N\to \infty$ limit. What happens for \emph{finite}
but large $N$? One would expect the sharp transition in Fig.
\ref{fig:stability} to be replaced by a smooth curve
(shown by the dashed lines in Fig. \ref{fig:stability}), but
can one describe analytically the precise form of this curve?
Also, is there any \emph{thermodynamical} sense to this 
stability-instability phase transition? 
If so, what is the analogue of \emph{free energy} and what is the
\emph{order} of this transition? Thanks to the recent developments 
in RMT on the statistics of $\lmax$ as reviewed in this article, 
it is possible to answer these
questions very precisely. In particular, we will see that
the large deviation function of $\lmax$ indeed plays the role of the 
\emph{free energy}
of an underlying Coulomb gas, with a third-order discontinuity
exactly at the critical point $\alpha=\alpha_c=1/\sqrt{2}$, thus
rendering it a {\it third order} phase transition. In addition,  
this third order phase transition turns out to be rather ubiquitous and occurs
in a wide variety of contexts. All these systems share a common
mechanism behind this third order phase transition that will be
elucidated in this article. 

The paper is organized as follows. In section 2, we summarize the main 
results for the statistics of $\lmax$ for large Gaussian random matrices, 
with a special focus on the large deviations. In section 3, we describe 
the Coulomb gas approach which provides a general framework to 
compute the left and right large deviation tails, which are separated by 
a third order phase transition. In section 4 we consider
various other systems where a similar third order transition occurs
and discuss, in some details, three cases namely
non-intersecting Brownian motions (and its relation to
$2$-d quantum chromodynamics), transport through a mesoscopic 
cavity and the entanglement entropy of a bipartite system
in a random pure state. In section 5, we describe the basic mechanism 
behind this third order phase transition and also discuss its higher 
order generalizations. Finally we conclude in section 6 with a summary
and discussion.

\section{Main results}
 
We consider $N \times N$ Gaussian random matrices with real symmetric, complex 
Hermitian, or  
quaternionic self-dual entries $X_{i,j}$ distributed via the joint law:
${\rm Pr}[\{X_{i,j}\}]\propto \exp\left[-c_0\, N\, {\rm Tr}(X^2)\right]$, where
$c_0$ is a constant. This distribution is
invariant respectively under orthogonal, unitary and symplectic
rotations giving rise to the three classical ensembles: Gaussian
orthogonal ensemble (GOE), Gaussian unitary ensemble (GUE) and
Gaussian symplectic ensemble (GSE). The eigenvalues 
and 
eigenvectors are consequently
random and their joint distributions decouple~\cite{Wig51,Meh91}. Integrating 
out the eigenvectors,
we focus here only on the statistics of $N$ eigenvalues  
$\lambda_1, \lambda_2, \cdots, \lambda_N$ which are all real. The joint 
probability density function (PDF) of these eigenvalues is given by the 
classical result~\cite{Wig51,Meh91,For10}
\begin{eqnarray}\label{joint_pdf0}
\hspace*{-1cm}P_{\rm joint}(\lambda_1, \lambda_2, \cdots, \lambda_N) =
B_N\, \exp\left[-c_0\, N\, \sum_{i=1}^N \lambda_i^2\right]\, \prod_{i<j} 
|\lambda_i-\lambda_j|^{\beta}\; ,
\end{eqnarray}  
where $B_N$ is a normalization constant and $\beta$ is called the Dyson index 
that takes quantized values
$\beta=1$ (GOE), $\beta=2$ (GUE) and $\beta=4$
(GSE). For convenience, we choose the constant $c_0=\beta/2$
and rewrite the statistical weight in (\ref{joint_pdf0}) as 
\begin{eqnarray}\label{joint_pdf}
\hspace*{-1cm}P_{\rm joint}(\lambda_1, \lambda_2, \cdots, \lambda_N) =
B_N(\beta) \exp\left[ -\beta\left( \frac{N}{2} \sum_{i=1}^N \lambda_i^2 
- \frac{1}{2}\sum_{i\neq j} \ln|\lambda_i - \lambda_j| \right) \right]\; .
\end{eqnarray}
Hence, this joint law can be interpreted as a Gibbs-Boltzmann 
measure~\cite{Dys62},
$P_{\rm joint}(\{\lambda_i\})\propto \exp\left[-\beta\, 
E\left(\{\lambda_i\}\right)\right]$,
of an interacting gas of charged particles on a line where $\lambda_i$ denotes
the position of the $i$-th charge and $\beta$ plays the role of the inverse 
temperature. The energy $E\left(\{\lambda_i\}\right)$ has
two parts: each pair of charges repel each other via a $2$-d Coulomb (logarithmic)
repulsion (even though the charges are confined on the $1$-d 
real line) 
and
each charge is subject to an external confining parabolic potential.
Note that while $\beta = 1$, $2$ and $4$ correspond to the three
classical rotationally invariant Gaussian ensembles, it is 
possible to associate a 
matrix model 
to (\ref{joint_pdf}) for any value of $\beta > 0$ (namely tridiagonal 
random matrices introduced in~\cite{DE02}). 
Here we focus on 
the largest eigenvalue $\lmax = \max_{1 \leq i \leq N} \lambda_i$: what 
can be said about its fluctuations, in particular when $N$ is large ? This 
is a nontrivial question as the interaction term, $\propto 
|\lambda_i - \lambda_j|^\beta$, renders inapplicable the classical results 
of extreme value statistics for 
i.~i.~d. random variables~\cite{Gum58}.

The two terms in the energy of the Coulomb gas in (\ref{joint_pdf}), the 
pairwise 
Coulomb repulsion and the external harmonic potential, compete with each other.
While the former tends to spread the charges apart, the later tends to confine
the charges near the origin. As a result of this competition, the system
of charges settle down into an equilibrium configuration on an average. 
One can estimate 
the typical value $\lambda_{\rm typ}$ of the eigenvalues by balancing
the two terms in the energy:
the potential 
energy, which is of order $\sim N^2\, {\lambda^2_{\rm typ}}$ and the 
interaction 
energy, which is of order $\sim N^2$: this yields $\lambda_{\rm typ} 
= {\cal O}(1)$. The average density of the charges is defined by
\begin{eqnarray}
\rho_N(\lambda)= \frac{1}{N}\, \left\langle \sum_{i=1}^N 
\delta(\lambda-\lambda_i)\right\rangle \;,
\label{avgden}
\end{eqnarray}
where the angular brackets denote an average with respect to the joint
PDF (\ref{joint_pdf}).
For such Gaussian matrices (\ref{joint_pdf}), it is well
known~\cite{Wig51,Meh91,For10} that as $N\to \infty$,  
the average density approaches an $N$-independent limiting form which
has a semi-circular shape 
on the compact support $[-\sqrt{2}, + \sqrt{2}]$ 
\begin{eqnarray}\label{semi_sc}
\lim_{N \to \infty} \rho_N(\lambda) = \tilde \rho_{\rm sc}(\lambda) = 
\frac{1}{\pi} \sqrt{2-\lambda^2} \;,
\end{eqnarray}
where $\tilde \rho_{\rm sc}(\lambda)$ is called the Wigner semi-circular 
law. Hence it follows from (\ref{semi_sc}) that the average location of 
$\lmax$ is given by the upper edge of the Wigner semi-circle:
\begin{eqnarray}\label{average_lmax}
\lim_{N \to \infty} \langle \lmax \rangle = \sqrt{2} \;.
\end{eqnarray}
From (\ref{rel_lmax},
\ref{stable_transition}), it follows that May's critical
point $1/\alpha_c=\sqrt{2}$ (see Fig. \ref{fig:stability}) coincides precisely 
with the
upper edge of the semi-circle, i.e., with $\langle \lmax\rangle=\sqrt{2}$.
However, for large but finite 
$N$, $\lmax$ will fluctuate from sample to sample 
and we would like to compute the full CDF of $\lmax$
\begin{eqnarray}
F_N(w) = {\rm Prob.} \, [\lmax \leq w] \;.
\end{eqnarray}

From the joint PDF in (\ref{joint_pdf}), one can express 
$F_N(w)$ as a ratio of two partition functions
\begin{eqnarray}\label{ratio_Z}
\hspace*{-1cm}&&F_N(w) = \frac{Z_N(w)}{Z_N(w \to \infty)} \;, \\
\hspace*{-1cm}&& Z_N(w) = \int_{-\infty}^w d \lambda_1 
\cdots \int_{-\infty}^w d \lambda_N 
\exp\left[ -\frac{\beta}{2}\left( N \sum_{i=1}^N \lambda_i^2 - 
\sum_{i\neq j} \ln|\lambda_i - \lambda_j| \right) \right] \;,
\end{eqnarray}
where $Z_N(w)$ has a clear physical interpretation: it is the partition 
function of a $2$-d Coulomb gas, confined on a $1$-d line and subject to a 
harmonic potential, in presence of a \emph{hard wall} at $w$~\cite{satyadean1}. 
The study of 
this ratio of two partition functions (\ref{ratio_Z}) reveals 
the existence of two distinct scales corresponding to (i) typical 
fluctuations of the top eigenvalue, where $\lmax = {\cal O}(N^{-2/3})$ and 
(ii) atypical large fluctuations where $\lmax = {\cal O}(1)$: these two 
cases need to be studied separately.
     
\subsection{Typical fluctuations}

To estimate the typical scale $\delta \lambda_{\max}$ of the fluctuations 
of $\lmax$, going beyond the estimate in (\ref{average_lmax}), one can 
apply the standard criterion of EVS, i.e.
\begin{eqnarray}
\int_{\sqrt{2} - \delta \lambda_{\max}}^{\sqrt{2}} 
\tilde \rho_{\rm sc}(\lambda) d \lambda \sim \frac{1}{N} \;,
\label{EVS.1}
\end{eqnarray}
which simply says that the fraction of eigenvalues to the right of the 
maximum (including itself) is typically $1/N$. Using the asymptotic 
behavior near the upper edge (\ref{semi_sc}), $\rho_{\rm 
sc}(\lambda) \propto (\sqrt{2}- \lambda)^{1/2}$ as $\lambda \to \sqrt{2}$,
one obtains~\cite{BB91,For93}
\begin{eqnarray}
\delta {\lmax} = \sqrt{2} - \lmax = {\cal O}(N^{-2/3})  \;.
\end{eqnarray}
More precisely, it turns out that as $N\to \infty$
\begin{eqnarray}
\lmax = \sqrt{2} + \frac{1}{\sqrt{2}}\, N^{-2/3}\, \chi_\beta \;,
\end{eqnarray}
where $\chi_\beta$ is an $N$-independent random variable. Its CDF, ${\cal F}_{\beta}(x)= {\rm Prob.}[\chi_\beta\le x]$, 
is 
known as the $\beta$-Tracy-Widom (TW) distribution which is known explicitly 
only for $\beta = 1,2$ and $4$. Tracy and Widom indeed obtained an 
explicit expression for $\beta = 2$ first \cite{TW94} and subsequently for 
$\beta = 1$ and $4$ \cite{TW96} in terms of the Hastings-McLeod solution 
of the Painlev\'e II equation
\begin{eqnarray}
q''(s) = 2 q^3(s) + s q(s) \;, \; q(s) 
\sim {\rm Ai}(s) \;, \, s \to \infty \;.
\end{eqnarray}
The CDF ${\cal F}_{\beta}(x)$ is then given explicitly for $\beta=1$, $2$
and $4$ by~\cite{TW94,TW96}
\begin{eqnarray}\label{TW_124}
&&{\cal F}_1(x) = \exp{\left[-\frac{1}{2} \int_x^\infty \left[ (s-x) q^2(s) + q(s) \right] \, ds\right]} \;, \\
&&{\cal F}_2(x) = \exp{\left[-\int_x^\infty  (s-x) q^2(s) \, ds\right]} \;, \nonumber \\
&&{\cal F}_4(2^{-\frac{2}{3}}x) = \exp{\left[-\frac{1}{2}  \int_x^\infty  (s-x) q^2(s) \, ds\right]} \cosh{\left[\frac{1}{2} \int_x^\infty q(s)\, ds \right]} \;. \nonumber
\end{eqnarray}
For other values of $\beta$ it can be shown that $\chi_\beta$ describes 
the fluctuations of the ground state of the following one-dimensional 
Schr\"odinger operator, called the ``stochastic Airy operator'' 
\cite{ES07, RRV11}
\begin{eqnarray}\label{stoc_Airy}
{\cal H}_{\beta} = -\frac{d^2}{dx^2} + x + \frac{2}{\sqrt{\beta}} \eta(x) \;,
\end{eqnarray}
where $\eta(x)$ is Gaussian white noise, of zero mean and with delta 
correlations, $\overline{\eta(x) \eta(x')} = \delta(x-x')$. For arbitrary 
$\beta>0$, the CDF ${\cal F}_\beta(x)$, or equivalently the
PDF ${\cal F}'_{\beta}(x)$ of $\chi_\beta$ has rather 
asymmetric 
non-Gaussian tails,
\begin{eqnarray}\label{asympt_TW}
{\cal F}'_\beta(x) \approx
\begin{cases}
&\exp{\left[-\dfrac{\beta}{24} |x|^3 \right]} \;, \; x \to - \infty \\
& \\
&\exp{\left[-\dfrac{2\beta}{3} x^{3/2} \right]} \;, \; x \to + \infty \;,
\end{cases}
\end{eqnarray}
where $\approx$ stands for a logarithmic equivalent [see below for more 
precise asymptotics (\ref{left_higher_order_TW}, 
\ref{right_higher_order_TW})]. These TW distributions also describe the 
top eigenvalue statistics of large real \cite{john,soscov} and complex 
\cite{johann} Gaussian Wishart matrices. Amazingly, the same TW 
distributions have emerged in a number of a priori unrelated problems 
\cite{maj} such as the longest increasing subsequence of random 
permutations \cite{baik}, directed polymers \cite{johann,poli} and growth 
models \cite{growth} in the Kardar-Parisi-Zhang (KPZ) universality class in 
$(1+1)$ 
dimensions as well as for the continuum $(1+1)$-dimensional KPZ 
equation~\cite{SS10,CLR10,DOT10,ACQ11}, 
sequence alignment problems 
\cite{sequence}, mesoscopic 
fluctuations in quantum dots \cite{dots}, height fluctuations of 
non-intersecting Brownian motions over a fixed time interval
\cite{FMS11,Lie12}, height fluctuations of non-intersecting interfaces
in presence of a long-range interaction induced by a 
substrate~\cite{NM_interface}, and also 
in finance 
\cite{biroli}. Remarkably, the TW distributions have been recently 
observed in experiments on nematic liquid crystals \cite{takeuchi} (for 
$\beta = 1,2$) and in experiments involving coupled fiber lasers 
\cite{davidson} (for $\beta = 1$).

\subsection{Atypical fluctuations and large deviations}     

\begin{figure}
\begin{center}
\includegraphics[width = 0.7\linewidth]{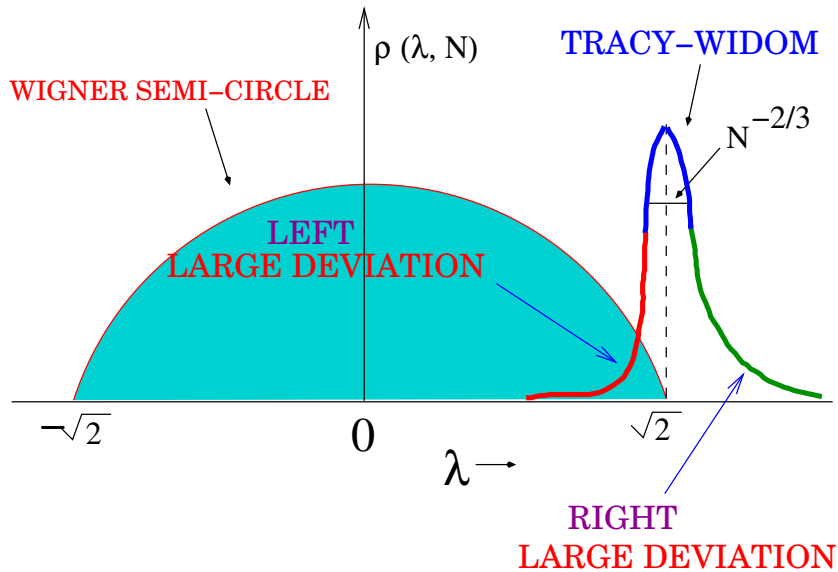}
\caption{Sketch of the pdf of $\lmax$ with a peak
around the right edge of the Wigner semicircle $\langle 
\lmax\rangle=\sqrt{2}$. The typical 
fluctuations of order $\mathcal{O}(N^{-2/3})$ around the mean
are described by the Tracy-Widom density (blue), while
the large deviations of order $\mathcal{O}(1)$ to the left and right
of the mean $\langle 
\lmax\rangle=\sqrt{2}$ are described by the left (red) and right (green)
large deviation tails.}
\label{fig:twld}
\end{center}  
\end{figure}

While the TW density describes the probability of \emph{typical} 
fluctuations
of $\lmax $ around its mean $\langle \lmax\rangle=\sqrt{2}$ on a 
\emph{small} scale
of $\sim \mathcal{O}(N^{-2/3})$, it does not describe   
\emph{atypically} large fluctuations, e.g. 
of order $\mathcal{O}(1)$ around its mean. Questions 
related to large deviations of extreme eigenvalues have recently emerged 
in a variety of contexts including cosmology~\cite{aazami,marsh1,marsh2}, 
disordered systems such as spin glasses 
~\cite{satyadean1,BrayDean2007,FyoWill2007,satyadean2, 
FN12,Fyo2013}, and in the assessment of the efficiency of 
data compression~\cite{majverg}. The probability of atypically large 
fluctuations, to leading order for large $N$, is described by two 
large deviations (or rate) functions $\Phi_{-}(x)$ (for fluctuations to 
the \emph{left} of the mean) and $\Phi_{+}(x)$ (for fluctuations to the 
\emph{right} of the mean). More precisely, the behavior of the CDF $F_N(w)$ of 
$\lmax$ for large but finite $N$ (as depicted schematically by the dashed lines 
in Fig. \ref{fig:stability}) is described as follows 
\begin{equation}\label{regimes_gaussian_CDF}
F_N(w)\approx
\begin{cases}
\exp\left[-\beta N^2 \Phi_{-}\left(w\right)\right]&,\,  
w<\sqrt{2} \; \& \; |w-\sqrt{2}|\sim\mathcal{O}(1)\\
\\
{\cal F}_\beta \left(\sqrt{2} 
N^{\frac{2}{3}}(w-\sqrt{2})\right)&, \, 
\hspace*{2.1cm} |w-\sqrt{2}|\sim\mathcal{O}(N^{-\frac{2}{3}})\\
\\
1-\exp\left[-\beta N \Phi_{+}\left(w\right)\right]&, 
\, w>\sqrt{2} \, \& \, |w-\sqrt{2}|\sim\mathcal{O}(1) \;.
\end{cases}
\end{equation}
Equivalently, the PDF of $\lmax$, obtained from the derivative $dF_N(w)/dw$
reads (keeping only leading order terms for large $N$)
\begin{equation}\label{regimes_gaussian}
{\cal P}(\lmax=w,N)\approx
\begin{cases}
\exp\left[-\beta N^2 \Phi_{-}\left(w\right)\right]&,\,  
w<\sqrt{2} \; \& \; |w-\sqrt{2}|\sim\mathcal{O}(1)\\
\\
\sqrt{2}N^{\frac{2}{3}}{\cal F}_\beta^\prime\left(\sqrt{2} 
N^{\frac{2}{3}}(w-\sqrt{2})\right)&, \, 
\hspace*{2.1cm} |w-\sqrt{2}|\sim\mathcal{O}(N^{-\frac{2}{3}})\\
\\
\exp\left[-\beta N \Phi_{+}\left(w\right)\right]&, 
\, w>\sqrt{2} \, \& \, |w-\sqrt{2}|\sim\mathcal{O}(1) \;.
\end{cases}
\end{equation}
A schematic picture of this probability density is shown in Fig. 
~\ref{fig:twld}.
We will see later that the physical mechanism responsible for the left 
tail ({\it pushed} Coulomb gas) is very different from the one on the 
right ({\it pulled} Coulomb gas).

Note that while the TW distribution ${\cal F}_\beta (x)$, describing the 
central part of the probability distribution of $\lambda_{\rm max}$, 
depends explicitly on $\beta$ [see Eq. (\ref{TW_124})], the two leading 
order rate functions $\Phi_{\mp}(w)$ are independent of $\beta$. 
Exploiting a simple physical method based on the Coulomb gas (see 
below), the left rate function $\Phi_{-}(z)$ was first explicitly computed 
in~\cite{satyadean1,satyadean2}
\begin{eqnarray}\label{phim}
\Phi_-(w) =&& \frac{1}{108}\Big[36w^2 - w^4 - (15w+w^3)\sqrt{w^2+6} \nonumber \\
&&+ 27 \left( \ln{18} - 2 \ln{\left(w + \sqrt{w^2+6}\right) } \right) \Big] \;, \; w < \sqrt{2}\;.
\end{eqnarray}
Note in particular the behavior when $w$ approaches the critical point 
from below:
\begin{eqnarray}\label{asympt_phim}
\Phi_-(w) \sim \frac{1}{6 \sqrt{2}} (\sqrt{2} - w)^3 \;, \; w \to \sqrt{2} \;.
\end{eqnarray}

On the other hand, the right rate function $\Phi_{+}(w)$
was computed in~\cite{majverg}. A more complicated, 
albeit mathematically rigorous, derivation (but only valid for $\beta=1$) of 
$\Phi_{+}(w)$ in the context of spin 
glass models can be found in \cite{ben}. Incidentally, the right tail of $\lmax$ can also be directly related
to the finite $N$ behavior of the average density of states to the right of the Wigner sea \cite{forrester_2012}. Indeed, for 
$\beta=1$, this finite $N$ right tail of the density was computed in Ref. \cite{Fyo2004}, from which one can extract the right rate function $\Phi_+(w)$. It reads
\begin{eqnarray}\label{phip}
\Phi_+(w) = \frac{1}{2} w \sqrt{w^2-2} + \ln{\left[ \frac{w - \sqrt{w^2-2}}{\sqrt{2}} \right]} \;,
\end{eqnarray}
with the asymptotic behavior
\begin{eqnarray}\label{asympt_phip}
\Phi_+(w) \sim \frac{2^{7/4}}{3} (w - \sqrt{2})^{3/2} \;, \; w {\to} \sqrt{2} \;.
\end{eqnarray}
More recently, the sub-leading corrections to the leading behavior have 
been explicitly computed using more sophisticated methods both for the 
left tail~\cite{BEMN}, as well as for the right tail~\cite{NM1,For12,BN1} [see 
also Eqs. (\ref{large_N_next}, \ref{right_higher_order}) below].
It is interesting to note that these explicit expressions for
the rate functions $\Phi_{\pm}(z)$ [respectively in Eqs. (\ref{phip})
and (\ref{phim})] have been used recently to compute exactly
the complexity of a class of spin glass models~\cite{FN12,Fyo2013}. Finally, we mention that a one-parameter extension of the rate
function $\Phi_+(w)$ in (\ref{phip}) was found in Ref. \cite{FLD13} in the context of the statistics of the global maximum of random quadratic forms over a sphere.

\subsection{Third order phase transition and matching}\label{section_matching}
  
The different behavior of ${\cal P}(\lmax = w, N) = F'_N(w)$ in  
(\ref{regimes_gaussian}) for $w < \sqrt{2}$ and $w > \sqrt{2}$ leads, in 
the limit $N \to \infty$, to a phase transition at the critical point $w_c 
= \sqrt{2}$. This is exactly the transition found by May in  
(\ref{stable_transition}), with $\alpha_c = 1/w_c = 1/\sqrt{2}$ (see Fig. 
\ref{fig:stability}). Here one can give a physical meaning to this 
transition as it corresponds to a thermodynamical phase transition for the 
free energy, $\propto \ln F_N(w)$, of a Coulomb gas, in presence of a wall 
(\ref{ratio_Z}) as the position of the wall $w$ crosses the critical value 
$w_c$. One has indeed, from~(\ref{regimes_gaussian_CDF}):
\begin{eqnarray}\label{transition_lmax}
\lim_{N \to \infty} - \frac{1}{N^2} \ln F_N(w) =
\begin{cases}
&\Phi_-(w)\;, \quad w < \sqrt{2} \;,\\
&0  \quad\quad\quad\quad w > \sqrt{2} \;,
\end{cases}
\end{eqnarray}   
where $\Phi_-(w)$ is given in (\ref{phim}). Since $\Phi_-(w) \sim 
(\sqrt{2} - w)^3$ when $w$ approaches $\sqrt{2}$ from below 
(\ref{asympt_phim}), the third derivative of the free energy of the 
Coulomb gas at the critical point $w_c = \sqrt{2}$ is discontinuous: this 
can thus be interpreted as a {\it third order phase transition}. 

This third order phase transition is very similar to the so called 
Gross-Witten-Wadia phase transition which was found in the 80's in the 
context of two-dimensional $U(N)$ lattice quantum chromodynamics (QCD) 
\cite{GW80,Wad80}. It was indeed shown in \cite{GW80,Wad80} that the
partition function $Z$ of the $U(N)$ 
lattice QCD in two dimensions with Wilson's action can be
reduced to $Z= \zeta^{N_p}$, where 
$N_p$ is the number of plaquettes in $2$-d and $\zeta= \int {\cal D}U 
\, \exp\left[(N/g)\, {\rm Tr} \left(U+U^{\dagger}\right)\right]$
is a single matrix integral. Here
$U$ is an $N\times N$ unitary matrix drawn from the uniform
Haar measure and $g$ is the coupling strength. By analyzing $\zeta$
in the large $N$ limit, it was shown that the free energy per plaquette
$ \lim_{N\to \infty} -(1/N^2)\ln \zeta$ undergoes
a third order phase transition at a critical coupling strength
$g_c=2$ separating a strong coupling phase ($g>g_c=2$)
and a weak coupling phase ($g<g_c=2$). In this case, 
because the matrix $U$ is unitary, 
its eigenvalues $\lambda_j$'s lie on the unit circle and 
are parameterized by angles $\theta_j$'s, 
with $\lambda_j = e^{i \theta_j}$. The average density of eigenvalues, 
in the limit $N \to \infty$, is given explicitly by \cite{GW80,Wad80}
\begin{eqnarray}
\label{GW_density}
\hspace*{-2cm}\rho^*{(\theta)} =
\begin{cases}
\dfrac{2}{\pi g} \cos{\left[\dfrac{\theta}{2} \sqrt{\dfrac{g}{2} - 
\sin^2\left(\dfrac{\theta}{2}\right)} \right]} \;, \; 0 \leq |\theta| 
\leq 2 \sin^{-1}\left(\sqrt{\dfrac{g}{2}} \right) &\;, \; {\rm for} \; g \leq {2} \\
\\
\dfrac{1}{2 \pi} \left(1 +\dfrac{2}{g} \cos \theta \right) \;, \; \theta \in [-\pi, +\pi] &\;, \;  {\rm for} \; g \geq {2} \;.
\end{cases}
\end{eqnarray}
When $g < 2$, the eigenvalues are thus confined on an arc of the circle
$-2 \sin^{-1}\left(\sqrt{\dfrac{g}{2}} \right)\le \theta\le 2 
\sin^{-1}\left(\sqrt{\dfrac{g}{2}} \right)$. As $g\to 2$ from below,
the charge density covers the full circle. Hence in this case the 
limiting angles $\pm 
\pi$ 
play the role of hard walls for the eigenvalues
and the gap between the edges of the arc and the hard wall $\pm 
\pi$ vanishes exactly as $g$ approaches the critical point $g=2$ from below. For $g>2$, the
density at the hard wall $\pm\pi$ has a nonzero finite value 
[see 
(\ref{GW_density})]. 

\begin{figure}\vspace*{1.0cm}
{ \includegraphics[height=5cm,width=5.8cm]{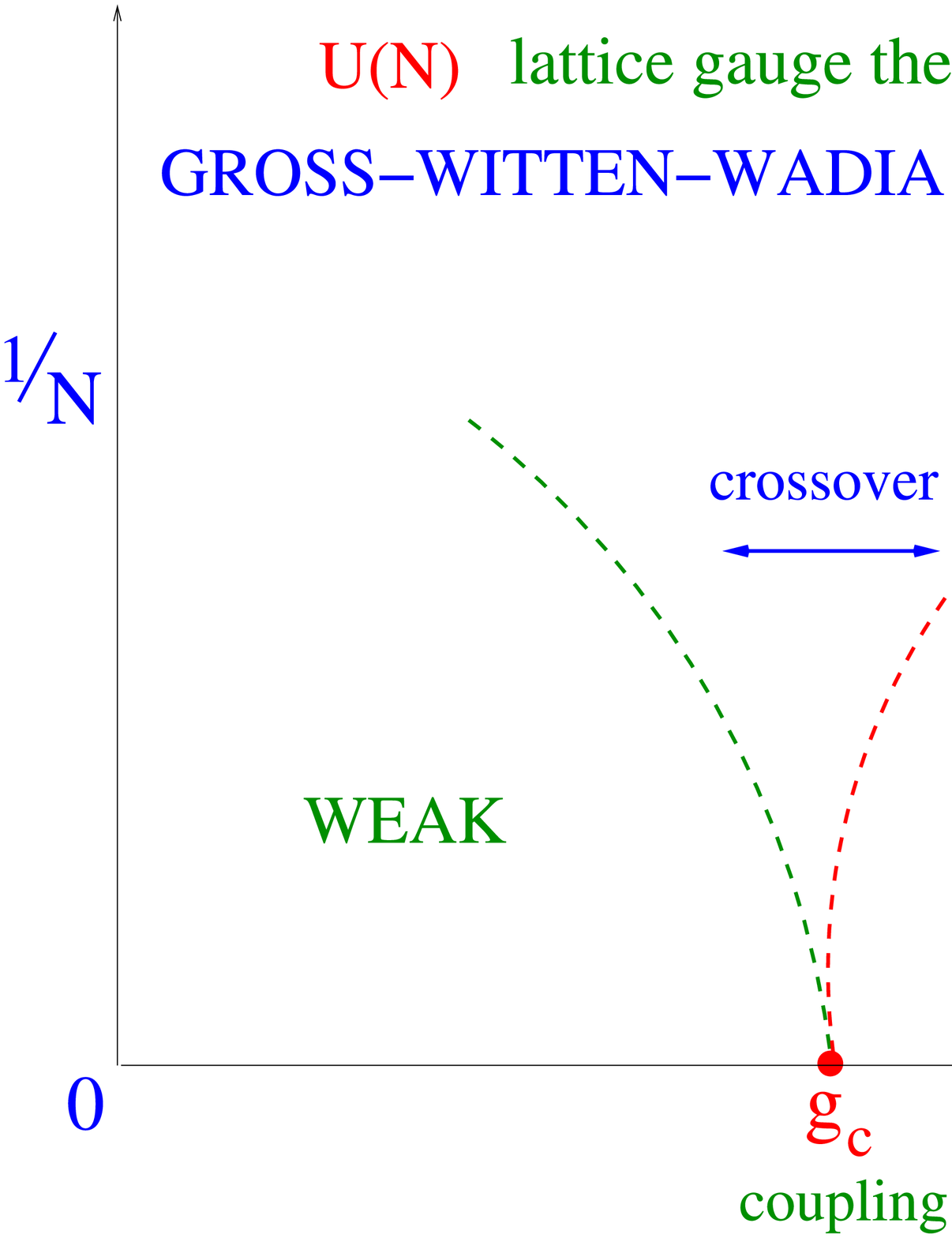}}
\hspace*{1.8cm}
{\includegraphics[height=5cm,width=5.8cm]{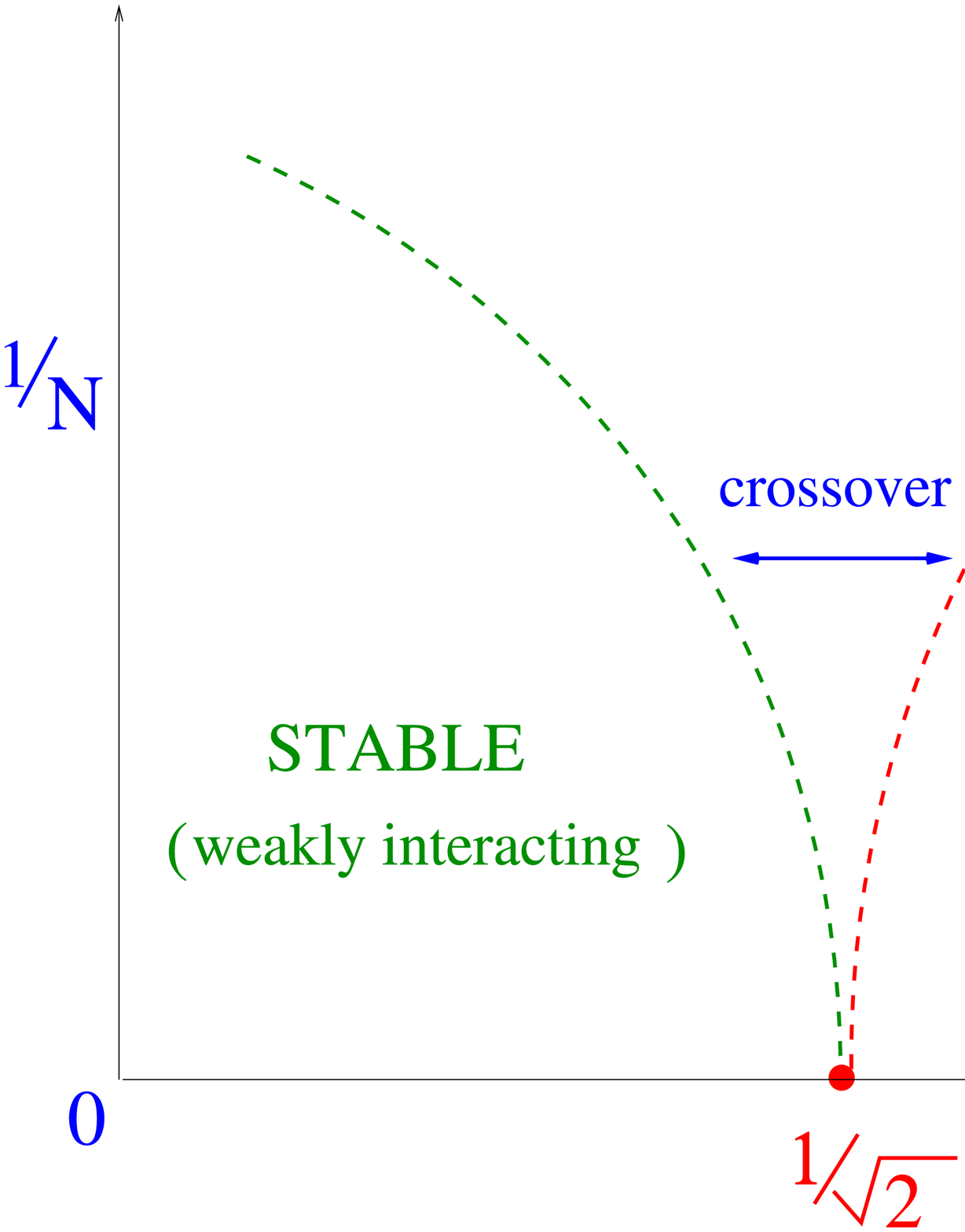}}
\caption{Phase diagrams of the $U(N)$ lattice QCD model
where $g$ is the coupling strength (left panel) and the
May's dynamical system with $\alpha$ denoting the interaction
strength between species (right panel).}
\label{fig:phdia}  
\end{figure}

This third order transition from
the strong to the weak coupling phase turns out
to be similar to the third order stable-unstable transition
in May's model as described in (\ref{transition_lmax})
(see also Fig. \ref{fig:stability}). Indeed, to bring out
the similarities between the two models one can draw
their respective phase diagrams as in Fig. 
\ref{fig:phdia}. Thus, the weak (strong) coupling phase in the $U(N)$ QCD 
is the 
analogue of the stable (unstable) phase in May's model.
In Fig. \ref{fig:phdia}, strictly in the $N\to \infty$ limit, one has
a sharp phase transition in both models as one goes through
the critical point on the horizontal axis. However, for finite but large 
$N$, this sharp phase transition gets rounded off and the `critical point'
gets splayed out into a {\it critical crossover zone}. As one increases $g$ or 
$\alpha$, the system crosses over
from the weak coupling (stable) to the strong coupling 
(unstable) phase over this critical zone. In May's model (right panel of Fig. 
\ref{fig:phdia}), while the large deviation functions in 
(\ref{regimes_gaussian}) describe the free energy deep inside the
two phases (stable and unstable), the Tracy-Widom  
distribution describes precisely the crossover behavior
of the free energy from one phase to the other as one traverses
the {\it critical zone} at finite $N$.
Later we will come back 
to a related third order transition, the so-called Douglas-Kazakov transition \cite{DK93}, found for continuum two dimensional QCD (see 
Fig.~\ref{fig_corres} below).


To investigate how smoothly this {\it crossover} occurs, 
it is interesting to match the central Tracy-Widom regime (around the peak)
in Fig. \ref{fig:twld} with the two far tails characterized 
by the rate functions in
(\ref{asympt_phim}) and (\ref{asympt_phip}). Indeed, one can check that the 
expansion of the large deviation functions around the critical point $w_c 
= \sqrt{2}$ matches smoothly with the asymptotic tails of the 
$\beta$-Tracy-Widom 
scaling function in the central region. To see this, let us first consider 
the 
left tail in (\ref{regimes_gaussian}), i .e. when $w <\sqrt{2}$. When 
$w \to \sqrt{2}$ from below we can substitute the asymptotic behavior of 
the rate function $\Phi_-(w)$ from (\ref{asympt_phim}) in the first line 
of (\ref{regimes_gaussian}). This yields for $1 \ll \sqrt{2} - w \ll 
\sqrt{2}$
\begin{eqnarray}\label{matching_left}
{\cal P}(\lmax=w,N)=\frac{d}{dw}F_N(w) \approx 
&&\exp{\left(-\frac{\beta}{6 \sqrt{2}}N^2(\sqrt{2}-w)^3 \right)} \;. 
\end{eqnarray}   
On the other hand, consider now the second line of  
(\ref{regimes_gaussian}) that describes the central typical fluctuations. 
When the deviation from the typical value $w_c = \sqrt{2}$ is large 
($\sqrt{2} - w \sim {\cal O}(1)$) we can substitute in the second line of 
(\ref{regimes_gaussian}) the left tail asymptotic behavior of the 
$\beta$-Tracy-Widom distribution (\ref{asympt_TW}) giving
\begin{eqnarray}
{\cal P}(\lmax=w,N)= \frac{d}{dw}F_N(w) \approx \exp{\left[-\frac{\beta}{24} 
\left[2^{1/2} N^{2/3}(\sqrt{2}-w)\right]^3\right]} \;,
\end{eqnarray}  
which after a trivial rearrangement, is identical to 
(\ref{matching_left}). This shows that the left tail of the central region 
matches smoothly with the left large deviation function. Similarly, on the 
right side, using the behavior of $\Phi_+(x)$ in (\ref{asympt_phip}), one 
finds from (\ref{regimes_gaussian}) that
\begin{eqnarray}
{\cal P}(\lmax=w,N)=\frac{d}{dw} F_N(w) \approx \exp\left(-\frac{2^{7/4} \beta}{3} N(w-\sqrt{2})^{3/2} \right) \;, 
\end{eqnarray}   
for $ 1 \ll w-\sqrt{2} \ll \sqrt{2}$, which matches with the 
right tail of the central part described by 
${\cal F}'_\beta(x)$ (\ref{asympt_TW}). 
Such a mechanism of matching between the central part and the 
large deviation tails of the distribution have been found in other 
similar problems \cite{majverg,VMB07} (see also Ref. \cite{SMCF} for a 
counterexample) that will be discussed later.

So far we have mainly focused on the Gaussian $\beta$-ensembles of random 
matrices, whose eigenvalues are distributed via (\ref{joint_pdf}). Other 
interesting ensembles of RMT include the Wishart random matrices~\cite{Wis28} 
(also 
called the Laguerre ensemble of RMT), which play an important role in 
Principal Component Analysis of large datasets. A Wishart matrix ${\bf W}$ 
can be viewed as a correlation matrix, built from the product ${\bf W = 
{\bf X}^\dagger {\bf X}}$ where ${\bf X}$ is a $M \times N$ random 
Gaussian matrix (real or complex). The joint PDF of the eigenvalues reads 
in this case~\cite{James64}
\begin{equation}\label{joint_wishart}
P^W_{\rm joint}(\lambda_1, \cdots, \lambda_N) = B^W_N(\beta,\gamma_0) 
\left(\prod_{i=1}^N \lambda_i^{\gamma_0\beta/2}\right) 
\Delta^\beta_N(\lambda_1,\cdots,\lambda_N) \, e^{-\frac{N\beta}{2}\sum_{i=1}^N \lambda_i} \;,
\end{equation}
where
\begin{equation}
\Delta_N(\lambda_1,\ldots, \lambda_N)=\prod_{1\leq i<j\leq N}(\lambda_i-\lambda_j) 
\label{vdm}
\end{equation}
is the Vandermonde determinant and $\gamma_0 = (1+M-N)-2/\beta$ and 
$B_N^W(\beta,\gamma_0)$ is a normalization constant. In this case, the 
Wigner semi-circle law (\ref{semi_sc}) for the average density
of eigenvalues is replaced by 
the Mar\v{c}enko-Pastur 
distribution \cite{MP67}. For $M\ge N$ and setting $c=N/M\leq 1$, 
the Mar\v{c}enko-Pastur density has a
compact support $[a,b]$ where $
a = (c^{-1/2} - 1)^2$ and $b = (c^{-1/2}+1)^2$ and is given explicitly by
\begin{eqnarray}\label{MP_law}
\lim_{N \to \infty} \frac{1}{N} \sum_{i=1}^N \langle \delta(\lambda-\lambda_i) \rangle = \tilde \rho_{\rm MP}(\lambda) = \frac{1}{2\pi \lambda} \sqrt{(\lambda-a)(b-\lambda)} \;.
\end{eqnarray}
Note that, like the Wigner semi-circle law (\ref{semi_sc}), $\tilde 
\rho_{\rm MP}(\lambda)$ vanishes like $\propto \sqrt{b-\lambda}$ near the 
right edge of the support. From (\ref{MP_law}) one deduces that $\lim_{N 
\to \infty}\langle \lmax \rangle = b$. While the typical 
fluctuations of $\lmax$, which are of order ${\cal O}(N^{-2/3})$, are also 
given by TW distributions \cite{john,soscov,johann}, the large deviations 
exhibit a behavior similar to, albeit different from  
(\ref{regimes_gaussian}):
\begin{eqnarray}
{\cal P}(\lmax = w,N) \approx
\begin{cases}
\exp{[-\beta N^2 \Psi_-(w)]} \;, \; &w < b \; \& \; |w-b| \sim {\cal O}(1) \;, \\
\\
\exp{[-\beta N \Psi_+(w)]}  \;, \; &w > b \; \& \; |w-b| \sim {\cal O}(1) \;,
\end{cases}
\end{eqnarray}
where the rate functions $\Psi_-(w)$ and $\Psi_+(w)$ have been computed 
exactly respectively in Ref. \cite{VMB07} and Ref.~\cite{majverg} (and are 
different from $\Phi_+(w)$ and $\Phi_-(w)$ found for Gaussian random 
matrices). Here also, using the behavior of $\Psi_-(w)$ when $w$ 
approaches the critical value $b$ from below, one can show that the CDF of 
$\lmax$ also exhibits a third order phase transition, similar to the one found 
for Gaussian $\beta$-ensembles (\ref{transition_lmax}). Remarkably, both 
large deviation functions $\Psi_-(w)$ and $\Psi_+(w)$ have been measured 
in experiments involving coupled fiber lasers \cite{davidson}.
We note that large deviation functions associated with
the minimum eigenvalue at the left edge of the Mar\v{c}enko-Pastur
sea (for $c<1$ strictly) have also been studied by similar Coulomb gas 
method~\cite{katzav_1,katzav_2}. For $c=1$ (the hard edge case where $M-N \ll 
\mathcal{O}(N)$ for large $N$), the
minimum eigenvalue distribution has also been studied
extensively~\cite{For93,Edelman_Wishart,For94} (for a recent review
see Ref.~\cite{Majumdar_review}), with
very nice applications in QCD~\cite{Verba_review} and in bipartite
quantum systems in a random pure state~\cite{MBL_2008}. Large deviations functions of $\lmax$ for
spiked Wishart ensembles were also computed in Ref. \cite{maida}. Finally, we point out that the large deviation functions
associated with Cauchy ensembles of random matrices have
recently been computed exactly using a similar Coulomb gas method~\cite{MSVV}.
  
\section{Derivation of large deviation tails using Coulomb gas method}
\label{section:CG}

In this section we briefly summarize the Coulomb gas method that
allows us to extract the large $N$ behavior of $F_N(w)={\rm Prob.}[\lmax\le 
w]$. We first express $F_N(w)$
as a ratio of two partition functions as in (\ref{ratio_Z}) 
where $Z_N(w)$ is expressed as a multiple $N$-fold integral
with a fixed upper bound at $w$
\begin{eqnarray}
&&Z_N(w) = \int_{-\infty}^w d \lambda_1 \cdots \int_{-\infty}^w d 
\lambda_N \exp{\left[-\beta N^2 E[\{{\lambda_i}\}] \right]} \label{multiple_int}\\
&& E[\{{\lambda_i}\}] = \frac{1}{2N} \sum_{i=1}^N \lambda_i^2 - 
\frac{1}{2N^2} \sum_{j \neq k} \log|\lambda_j - \lambda_k| \;.
\end{eqnarray} 
The main idea then is to evaluate this multiple integral for large $N$, but 
with a fixed $w$,
via the steepest descent (saddle point) method.
When $N$ is large, the Coulomb gas with $N$ discrete charges can be
characterized by a continuous charge density 
\begin{eqnarray}
\rho_w(\lambda) = \frac{1}{N} \sum_{i=1}^N \delta(\lambda-\lambda_i) \;,
\end{eqnarray}
such that $\rho_w(\lambda) d\lambda$ counts 
the fraction of eigenvalues between $\lambda$ and $\lambda+ d \lambda$. It 
is normalized to unity and, because of the presence of the wall, one has 
obviously 
$\rho_w(\lambda) = 0$ for $\lambda > w$. The next step is then to replace 
the multiple integral in (\ref{multiple_int}) by a functional integral 
over the space of all possible normalized densities $\rho_w(\lambda)$. 
This gives~\cite{satyadean1,satyadean2}
\begin{eqnarray}
&&Z_N(w) \propto \int {\cal D}[\rho_w] \exp{\left[-\beta N^2 {\cal 
E}[\rho_w] + {\cal O}(N)\right]}\, \delta\left(\int_{-\infty}^w 
\rho_w(\lambda)d\lambda-1\right) \;, \label{functional_int}\\
&& {\cal E}[\rho_w] = 
\frac{1}{2} \int_{-\infty}^w \lambda^2 \rho_w(\lambda) - 
\frac{1}{2} \int_{-\infty}^w d \lambda \int_{-\infty}^w d 
\lambda' \rho_w(\lambda) \rho_w(\lambda') \ln(|\lambda- \lambda'|) \;,
\end{eqnarray}
where the terms of order ${\cal O}(N)$ in the exponent in  
(\ref{functional_int}) come from the entropy term associated with the 
density 
field $\rho_w$ when going from the multiple $N$-fold integral
to a functional integral~\cite{Dys62}. Roughly speaking, it corresponds 
to all microscopic charge configurations compatible with
a given macroscopic density $\rho_w(\lambda)$. This entropic 
contribution was explicitly computed recently  
for Gaussian ensembles 
in Ref.~\cite{ABG12}
and for the Wishart-Laguerre 
ensembles in Ref.~\cite{ABMV13}. But here, we will mainly be concerned
with the leading energy term $\sim \mathcal{O}(N^2)$ and hence will
ignore the entropy term.

We next evaluate the functional integral in the large $N$ limit using the
saddle point method. The 
density at the saddle point $\rho^*_w(\lambda)$ minimizes the energy ${\cal
E}[\rho_w]$ subject to the constraint $\int_{-\infty}^w \rho^*_w(\lambda)
\, d \lambda = 1$. Hence $\rho^*_w(\lambda)$ is a stationary point of the 
following
action $S[\rho_w (\lambda)]$
\begin{eqnarray}\label{saddle_action}
\frac{\delta S[\rho_w]}{\delta \rho_w} \Bigg
|_{\rho_w = \rho^*_w} = 0 \;, \; S[\rho_w] =
{\cal E}[\rho_w] + C\left(\int_{-\infty}^w \rho_w(\lambda)
d\lambda  - 1 \right) \;,
\end{eqnarray}
where $C$ is a Lagrange multiplier that ensures the normalization 
condition of the density. Another alternative way to arrive
at the same result is to replace the delta function in  
(\ref{functional_int}) by its integral representation and then
minimize the resulting action. Once the saddle point density 
$\rho^*_w(\lambda)$ is
found from (\ref{saddle_action}), one can compute the CDF of $\lmax$  
from (\ref{ratio_Z}, \ref{functional_int}) as
\begin{eqnarray}\label{rel_phim_energy}
F_N(w) \approx \exp{\left[- \beta N^2\, \left({\cal E}[\rho^*_w] - {\cal 
E}[\rho^*_\infty]\right)\right]} \;,
\end{eqnarray}
where $\rho^*_\infty(\lambda) = \lim_{w \to \infty} \rho^*_w (\lambda)$.
Thus we see that the cumulative distribution of $\lmax$ (which is a
probabilistic quantity) can be interpreted thermodynamically. Its logarithm
can be expressed as the free energy difference between two Coulomb 
gases: one in presence of a hard wall at $w$ and the other is free, i.e.,
the wall is located at infinity.

The next step is thus to determine the solution
of the saddle point equation (\ref{saddle_action}).
Setting the functional derivative to zero in (\ref{saddle_action})
gives the following integral equation
\begin{eqnarray}\label{saddle_0}
\frac{\lambda^2}{2} - \int \rho^*_w(\lambda') 
\ln(|\lambda - \lambda'|) d \lambda' + C = 0 \;,
\end{eqnarray}
which is valid only over the support of $\rho^*_w(\lambda)$, i.e, where this
density is nonzero. 
Clearly, the solution can not have an unbounded
support. Because, if that was so, for large $\lambda$, the first term in
(\ref{saddle_0}) grows as $\lambda^2$ whereas the second term grows as $\ln(|\lambda|)$
and hence they can never balance each other. Evidently, then the solution
must have a finite support over $[a_1,a_2]$ and assuming that this is
a single compact support, the range of integration in (\ref{saddle_0})
can be set from $a_1$ to $a_2$ (with $a_2>a_1$). Deriving once again
(\ref{saddle_0}) with respect to $\lambda$ (and for $\lambda \in [a_1,a_2]$),
we can get rid of the constant $C$ and get a singular Cauchy type equation
\begin{eqnarray}\label{saddle_3}
\lambda =  \dashint_{a_1}^{a_2} \frac{\rho_w^*(\lambda')}{\lambda - \lambda'}  \;,
\, d \lambda' \; 
\end{eqnarray} 
where $\dashint$ denotes the principal value of the integral.
Eq. (\ref{saddle_3}) belongs to the general class of Cauchy singular
integral equations (with one compact support) of the form
\begin{eqnarray}
g(\lambda)= \dashint_{a_1}^{a_2} \frac{\rho(\lambda')}{\lambda - \lambda'}
\, d \lambda' \; , 
\label{cauchy.gen.1}
\end{eqnarray}
valid over the single support $\lambda\in [a_1,a_2]$ and with an arbitrary 
source 
function $g(\lambda)$.  The problem is to invert this equation, i.e.,
find $\rho(\lambda)$ given $g(\lambda)$.
Fortunately, such singular integral equations can be 
explicitly inverted using a formula due to 
Tricomi~\cite{Tri85} that reads
\begin{eqnarray}
\rho(\lambda)= \frac{1}{\pi 
\sqrt{(a_2-\lambda)(\lambda-a_1)}}\left[C_0-\dashint_{a_1}^{a_2} \frac{dt}{\pi} 
\frac{\sqrt{(a_2-t)(t-a_1)}}{\lambda-t}\, g(t)\right]
\label{tricomi_sol_gen}
\end{eqnarray}
where $C_0= \int_{a_1}^{a_2} \rho(\lambda) d\lambda$ is a constant. 
In our case, the source function $g(\lambda)=\lambda$ and the 
constant $C_0=1$ due to the normalization $\int_{a_1}^{a_2} \rho_w(\lambda) 
d\lambda=1$.
Fortunately, the principal value of the integral in 
(\ref{tricomi_sol_gen}) with $g(t)=t$ can be explicitly 
computed. The unknown edges $a_1$ and $a_2$ can also
be completely determined leading to the following
exact result~\cite{satyadean1,satyadean2}
\begin{eqnarray}\hspace*{-2cm}
\rho^*_w(\lambda) =
\begin{cases}
\dfrac{1}{\pi} \sqrt{2-\lambda^2},\quad {\rm with}\quad 
-\sqrt{2}\leq \lambda \leq \sqrt{2} \quad &{\rm for}\;\; w>\sqrt{2} \\
\\
\dfrac{\sqrt{\lambda+L(w)}}{2\pi \sqrt{w-\lambda}}
\left[w + L(w)-2\lambda\right] \quad {\rm with} \quad  -L(w) \leq 
\lambda
\leq w  \quad &{\rm for}\;\; w < \sqrt{2} \label{rhostar}
\label{wigner.1} 
\end{cases}
\end{eqnarray}
where
\begin{eqnarray}\hspace*{-1.5cm}
L(w) = \frac{2\sqrt{w^2+6} - w}{3}\, .
\label{lw}
\end{eqnarray}
Note in particular that, in presence of a pushing wall ($w<\sqrt{2}$), the 
density  
diverges close to the wall,
$\rho^*_w(\lambda) \propto
1/\sqrt{w-\lambda} $ as $\lambda\to w$.
\begin{figure}
\begin{center}
\includegraphics[width=\linewidth]{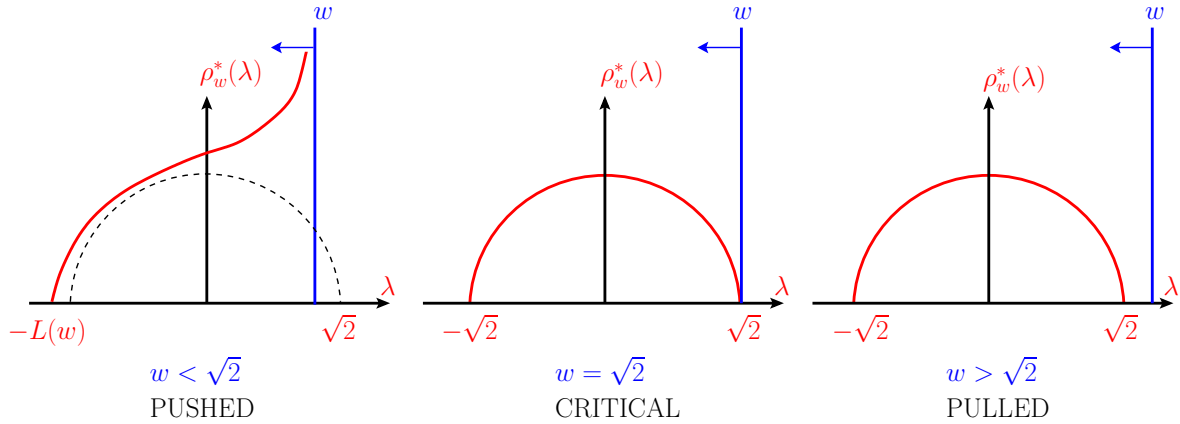}   
\caption{Effect of the presence of a wall on the Wigner semi-circle.
If $w<\sqrt{2}$, the density is {\it pushed} which leads to a
complete reorganization of the density charges will if
$w > \sqrt{2}$ a single charge is {\it pulled}, leaving the bulk of
the density unchanged.}\label{fig_wall}
\end{center}
\end{figure}

Thus, to leading order for large $N$, the saddle density sticks to the 
Wigner semi-circular form
as long as $w>w_c=\sqrt{2}$, but changes its form when $w<w_c=\sqrt{2}$ (see
Fig. \ref{fig_wall}).
This   
solution makes complete physical sense. Consider first the limit $w\to \infty$.
In this case, the equilibrium charge density of the Coulomb
gas is evidently of the Wigner semi-circular form. Now, imagine bringing 
the wall from infinity closer to the origin. As long as the wall position
is bigger than the right edge $\sqrt{2}$ of the semi-circle, the charges
do not feel the existence of the wall and are happy to equilibrate into
the semi-circular form, thus giving rise to the solution in (\ref{wigner.1})
for $w>\sqrt{2}$. When the wall position $w$ hits the edge of the
semi-circle, i.e., when $w\to \sqrt{2}$ from above, the charges start feeling 
the wall
and finally when $w<\sqrt{2}$, i.e., the charge density gets {\em pushed} by 
the wall,
the charges have to re-organize to find a new equilibrium density that
minimizes the energy, leading to the new deformed solution in  
(\ref{rhostar}). This is indeed the mechanism behind the phase transition,
driven by the vanishing of the the gap between the wall and the edge of the Coulomb 
gas droplet.

Finally, injecting the saddle point solution into the action in  
(\ref{rel_phim_energy}), one then gets, to leading order for large $N$, the result announced in (\ref{transition_lmax})
\begin{eqnarray}
F_N(w) \approx 
\begin{cases}
\exp\left[-\beta N^2 \Phi_{-}(w)\right]\;\; &{\rm for}\;\; 
w<\sqrt{2} \;, \label{free.left} \\
\\
1 \hspace*{3.5cm} & {\rm for}\;\; w>\sqrt{2} \;,\label{free.right}
\end{cases}
\end{eqnarray}
where $\Phi_{-}(w)$ is given explicitly in (\ref{phim}).
Note that the complete  
re-organization of the charge density for $w<w_c$ costs an energy of order
$\sim {\cal O}(N^2)$, and $\Phi_{-}(w)$ is proportional to this energy cost.
The left rate function $\Phi_{-}(w)$
vanishes
as $\sim (\sqrt{2}-w)^3$ as $w$ approaches to the critical point $\sqrt{2}$ from 
below, thus making the transition third order. 

As an aside, we note that recently the higher order corrections for the 
left tail were 
computed in \cite{BEMN} using a method based on the so called 
loop-equations an their large $N$ expansion \cite{CEM11}. It was shown 
that
\begin{eqnarray}\label{large_N_next}
- \ln \left[ {\cal P}(\lmax=w,N)\right] = 
&&N^2 \Phi_+(w) + N (\beta-2) \Psi_1(w) + (\ln N) \phi_{\beta}(w) \nonumber \\
&&+ \Psi_2(\beta,w) + {\cal O}(1/N) \;,  
\end{eqnarray}
where the functions $\Psi_1, \phi_\beta$ and $\Psi_2$ were computed 
exactly. By matching the left tail with the central part, described by the 
$\beta$-TW distribution (\ref{regimes_gaussian}), it is possible -- as 
discussed above -- to deduce from (\ref{large_N_next}) the higher order 
asymptotic expansion of ${\cal F}'_\beta$ as \cite{BEMN}
\begin{eqnarray}\label{left_higher_order_TW}
{\cal F}'_\beta(x) \underset{x\to -\infty}{\sim} \tau_\beta 
|x|^{\frac{\beta^2+4-6\beta}{16\beta}}
\exp{\left[-\beta \frac{|x|^3}{24} + \sqrt{2} \frac{\beta-2}{6} |x|^{3/2}\right]} \;,
\end{eqnarray}
which generalizes the result of Ref. \cite{BBdF08} valid for $\beta = 1,2$ 
and $4$ to any real value of $\beta >0$ (including the constant 
$\tau_\beta$ which can be computed explicitly for any real value of 
$\beta$~\cite{BEMN}).

\vspace{1cm}

\noindent{\it Right large deviation tail} ($w > \sqrt{2}$): The leading order large $N$ saddle point solution in the
previous subsection yields a nontrivial
left rate function $\Phi_{-}(w)$ associated with $F_N(w)$ 
(\ref{free.left}) for $w<\sqrt{2}$, corresponding to the {\it unstable} phase
of May's model, but 
provides only a trivial answer 
$F_N(w)\sim 1$ for $w>\sqrt{2}$ (i.e., in the {\it stable} phase of May's 
model).
This is actually very similar to the QCD model in $2$-d, where it is
known~\cite{GW80,GM94} that the saddle point solution gives nontrivial 
$1/N$ corrections to the free 
energy
only in the strong coupling phase (analogue of the unstable phase), while 
it gives a trivial result in the weak coupling phase (analogue of 
the
stable phase). The deep reason for this is that in the weak coupling
phase the corrections to the free energy are essentially {\it non-perturbative}
that can not be captured via an $1/N$ expansion of the free energy~\cite{GM94}.
In gauge theory, these non-perturbative corrections correspond
to instanton solutions~\cite{GM94,Marino}. In the present case, to capture 
the 
nontrivial
non-perturbative corrections in the stable phase ($w>\sqrt{2}$) and go beyond 
the 
trivial lowest order result $F_N(w)\sim 1$, we 
need to find a similar ``instanton-like'' strategy which is outlined below. 
   
Following \cite{majverg}, the right strategy turns out to 
consider directly the PDF of $\lmax$, rather than its CDF $F_N(w)$. Taking 
derivative of  
(\ref{ratio_Z}) with respect to $w$ yields an exact expression for the 
PDF of $\lmax$, 
\begin{equation}\label{pdf_right}
{\cal P}(\lmax=w,N) \propto e^{- N \beta \frac{w^2}{2}}
\int_{-\infty}^w d \lambda_1 \cdots \int_{-\infty}^w d \lambda_{N-1} \; 
e^{\beta \sum_{j=1}^{N-1} \ln{(|w-\lambda_j|)}} 
P_{\rm joint} (\lambda_1, \cdots, \lambda_{N-1}) \;,
\end{equation} 
where $P_{\rm joint}(\lambda_1, \cdots, \lambda_{N-1})$ is the joint PDF 
given in (\ref{joint_pdf}) for $(N-1)$ eigenvalues. The idea
then is to evaluate this $(N-1)$-fold integral via the saddle
point method, for a fixed $w-\sqrt{2} \sim {\cal O}(1)$. In this case,
one expects that only one (out of a large number $N$) charge at $w-\sqrt{2}\sim 
{\cal O}(1)$ does not disturb (to leading order) the equilibrium
configuration of the rest $(N-1)$ charges whose density then still
remains of the standard semi-circular form (\ref{wigner.1})
(see Fig. \ref{fig_wall}). Following this physical picture,
the multiple $(N-1)$-fold integral in (\ref{pdf_right}) is then well 
approximated
by $\langle e^{\beta\sum_{j=1}^{N-1} \ln (w-\lambda_j)}\rangle$ where
the angular brackets denote an average evaluated at the saddle point
with semi-circular density. To leading order in large $N$, one
can further replace the average of the exponential by the
exponential of the average $e^{\langle 
\beta\sum_{j=1}^{N-1} \ln 
(w-\lambda_j)\rangle}$. This gives~\cite{majverg} 
\begin{eqnarray}
\label{pdf_right_1}
{\cal P}(\lmax=w,N)\propto \exp\left[-\beta N \frac{{w}^2}{2}\,
+\beta N
\int
\ln |{w}-\lambda|\, {\tilde \rho}_{\rm sc}(\lambda)\,d\lambda\right]\; ,
\end{eqnarray}
where ${\tilde \rho}_{\rm sc}(\lambda)= (1/\pi)\sqrt{2-\lambda^2}$.
Thus, one gets to leading order for large $N$
\begin{eqnarray}
{\cal P}(\lmax=w,N)\sim \exp\left[-\beta N \Phi_+(w)\right] \;,
\label{pdf_right_2}
\end{eqnarray}
where the right rate function $\Phi_+(w)$ is given by, up to an overall
normalization constant
\begin{equation}\label{pdf_rate}
\Phi_+(w)  =  \frac{w^2}{2} - 
\int_{-\sqrt{2}}^{\sqrt{2}} 
\ln{(|w - \lambda|) \rho_{\rm sc}(\lambda)} d\lambda  + A \; , \quad\quad w > 
\sqrt{2} \;,
\end{equation}
where the constant $A$ is computed such that $\Phi_+(w = \sqrt{2}) = 0$, 
since our reference configuration is the one where $\lmax = \sqrt{2}$. 
Evaluating the integral in (\ref{pdf_rate}), one obtains the result for 
$\Phi_+(w)$ given in (\ref{phip}).

\begin{figure}\vspace*{3cm}
\begin{center}
\includegraphics[width = 0.7\linewidth]{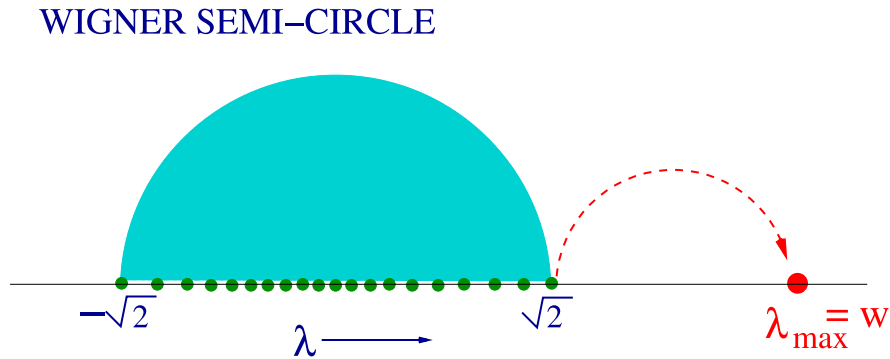}
\caption{The right rate function is evaluated by computing the
energy cost in {\it pulling} a single charge at $w-\sqrt{2}\sim O(1)$ out of 
the Wigner sea
to its right.}
\label{fig:pulled}
\end{center}
\end{figure}

Thus physically, the quantity $\beta N \Phi_+(w)$ in the right tail of the PDF 
of $\lmax$ just corresponds to the energy cost $\Delta E$ in {\it pulling}
the rightmost charge out of the Wigner sea (see Fig. \ref{fig:pulled}). Since 
only one charge
goes out of the Wigner sea (and it does not lead to a re-organization of the
rest of $(N-1)$ charges as in the case $w<\sqrt{2}$),
the energy cost $\Delta E$ is of order ${\cal O}(N)$
and is estimated in (\ref{pdf_rate}) by computing the energy of the 
rightmost charge
in the external quadratic potential and its Coulomb interaction energy
with the rest of the Wigner sea.  

To compute the 
higher order corrections to the right tail one needs more sophisticated 
techniques. These were obtained in 
\cite{NM1} for $\beta = 2$, using a method based on orthogonal 
polynomials over the unusual interval $(-\infty, w]$ and adapting a technique 
originally developed in the context of 
QCD \cite{GM94}. It was found \cite{NM1} that
\begin{eqnarray}\label{right_higher_order}
\frac{d}{dw} F_N(w) {\sim} \frac{e^{-2 N \Phi_+(w)}}{2\pi \sqrt{2}(w^2-2)} 
\;, \; w > \sqrt{2} \;,
\end{eqnarray} 
which, by matching with the central part (\ref{regimes_gaussian}), yields 
the asymptotic behavior of ${\cal F}'_2(x)$ beyond the leading order given 
in (\ref{asympt_TW}). One can show that it agrees with the rigorous result found in 
Ref. \cite{DV11} for the right tail of the $\beta$-TW distribution:
\begin{eqnarray}\label{right_higher_order_TW}
1 - {\cal F}_{\beta}(x) = x^{-\frac{3\beta}{4} + o(1)} {e^{-\frac{2}{3} 
\beta x^{3/2}}} \;,
\end{eqnarray} 
which was obtained using the stochastic Airy operator representation 
(\ref{stoc_Airy}). Very recently, the right large
deviation behavior of $\lambda_{\rm max}$ has been computed
to all orders in $N$ by a generalized loop equation method by Borot and 
Nadal~\cite{BN1} (see also Ref.~\cite{For12}).   
Finally, the unusual orthogonal polynomial method developed
in Ref. \cite{NM1} has recently been extended and generalized to 
matrix models with higher order critical 
points~\cite{Akemann_Atkin,Atkin_Zohren}.

\section{Third order phase transitions in other physical models} 

The third order phase transition, discussed in detail above for $F_N(w)$ in 
the context of the top eigenvalue of Gaussian random matrices 
(\ref{transition_lmax}), also occurs in various other contexts. We have 
already mentioned that this transition is very similar to the one found in 
$2$-d lattice QCD, the so called Gross-Witten-Wadia transition 
~\cite{GW80,Wad80}. It also appears in the continuum QCD model
in two dimensions--the so called Douglas-Kazakov transition~\cite{DK93}.
Recently, similar third order transition has been found
in the large deviation function associated with  
the distribution of the
maximum height of a set of non-intersecting Brownian excursions
in one dimension~\cite{FMS11,SMCF}, the 
distribution
of conductance through mesoscopic cavities~\cite{VMB08,VMB10,DMTV11}
and the distribution of Renyi entanglement entropy in a bipartite random pure
state~\cite{NMV10,NMV11} (see also Refs.~\cite{Facchi08,Facchi10} for
a slightly different description of the same physical system in terms of the 
Laplace
transform of the distribution of purity)--these three 
cases will be discussed in some detail
in this section. In addition, similar third order phase transitions have been 
noted
in models of information propagation through multiple input multiple output
(MIMO) channels~\cite{MIMO}, in the behavior of the complexity
in simple spin glass models~\cite{FN12}, and more recently in 
the combinatorial
problem of random tilings~\cite{CP2013}. 
We 
will 
see later that
all these different problems share a common mechanism behind the
third order phase transition-- 
it happens when the gap, between the 
soft edge of the Coulomb charge density (supported over a sinhle interval) 
with a square-root singularity at its edge and a hard 
wall, vanishes as a control parameter (for instance the coupling strength 
$\alpha$ in May's model (\ref{linear_stability}) or the gauge field coupling
$g$ in 
$2$-d lattice 
QCD) crosses a critical value. 

\subsection{Maximal height of $N$ non-intersecting Brownian excursions}
 
\begin{figure}[hh]
\begin{center}
\includegraphics[width=0.6\linewidth]{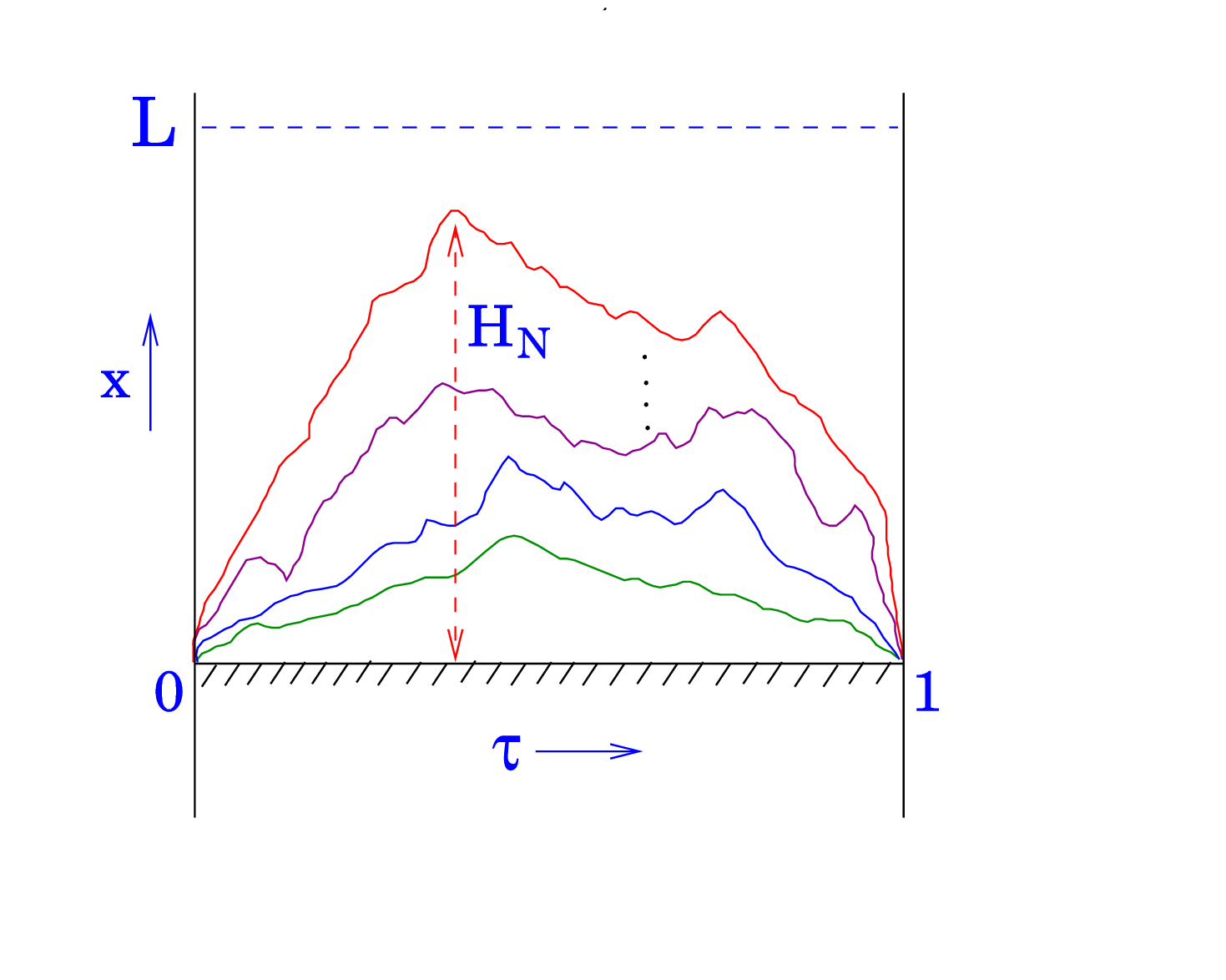}
\caption{Trajectories of $N$ non-intersecting Brownian motions 
$x_1(\tau)<x_2(\tau)<\ldots < x_N(\tau)$, all start at the origin
and return to the origin at $\tau=1$, staying positive in between.
$\tilde{F}_N(L)$ denotes the probability
that the maximal height $H_N=\max_\tau\{x_N(\tau), 0\le \tau\le 1\}$ stays 
below the level $L$ over the time interval $0\le \tau\le 1$.}
\label{fig:watermelon}
\end{center}
\end{figure}

We consider $N$ non-intersecting Brownian motions on a line, $x_1(\tau) < 
x_2(\tau) 
\cdots < x_N(\tau)$ with an absorbing boundary condition in $x=0$. In 
addition, the walkers start in the vicinity of the origin and are 
conditioned to return to the origin exactly at $\tau=1$ (see Fig. 
\ref{fig:watermelon}). Such configurations of Brownian motions are called 
non-intersecting Brownian excursions \cite{TW07}, or sometimes 
``watermelons with a wall'' \cite{BM03}. An interesting observable is the 
so called height of the watermelon, which has been studied in the past 
years by several authors \cite{FMS11,Lie12,BM03,SMCR,KIK08,KK08,Fei12,RS10,RS11} 
(see also \cite{BFPSW09} for a related quantity in the context of Dyson's 
Brownian motion).

\begin{figure}[ht]
\begin{center}
 \includegraphics[width = \linewidth]{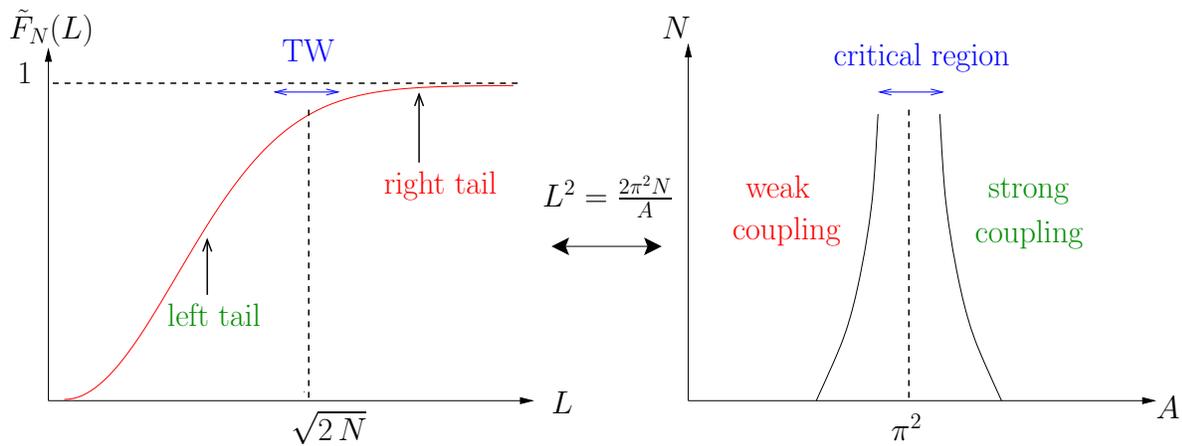}
\caption{{\bf Left:} 
Schematic sketch of the CDF of $H_N$, $\tilde F_N(L)$ as defined in  
(\ref{Fp}) for $N$ vicious walkers on the line segment $[0,L]$ with an 
absorbing boundary condition in $x=0$. {\bf Right:} Sketch of the phase 
diagram in the plane $(A,N)$ of two-dimensional Yang-Mills theory on a 
sphere with the gauge group ${\rm Sp}(2N)$ as obtained in Ref. 
\cite{DK93,CNS96}. The weak (strong) coupling phase in the right panel 
corresponds to the right (left) tail of $\tilde F_N(L)$ in the left panel. 
The critical region around $A=\pi^2$ in the right panel corresponds to the 
Tracy-Widom (TW) regime in the left panel around the critical point 
$L_c(N)= \sqrt{2N}$.}
\label{fig_corres}
\end{center}
\end{figure}

The height $H_N$ of the watermelon is defined as the maximal displacement 
of the topmost path in this (half-) watermelon configuration, i.e.
\begin{eqnarray}
H_N = {\max}_{\tau} [x_N(\tau), \tau \in [0,1]] \;.
\end{eqnarray} 
The CDF of $H_N$ was computed exactly in Ref. \cite{SMCR} using a 
Fermionic path integral (see also \cite{KIK08,Fei12} for related 
computations using different methods), yielding the result
\begin{align}\label{Fp}
\tilde{F}_N(L)&:= {\rm Prob.}\,[H_N\leq L] \nonumber \\
&=\frac{A_N}{L^{2N^2+N}}\sum_{n_1=-\infty}^\infty \ldots \sum_{n_N=-\infty}^{\infty} 
\Delta^2(n_1^2,\ldots,n_N^2)\Big(\prod_{j=1}^N n_j^2\Big)
e^{-\frac{\pi^2}{2L^2}\sum_{j=1}^N 
n_j^2} \;,
\end{align}
where $\Delta_N(y_1,\ldots, y_N)$ is the Vandermonde determinant 
(\ref{vdm}) and where $A_N$ is a normalization constant. In Ref. 
\cite{FMS11} it was shown that this CDF in the Brownian motion model 
(\ref{Fp}) maps onto the exactly solvable partition function (up to a 
multiplicative pre-factor) of a two-dimensional Yang-Mills gauge theory. More 
precisely, if one denotes~by ${\cal Z}(A,G)$ the partition function of the 
two-dimensional (continuum) Yang-Mills theory on the sphere (denoted as 
$YM_2$) with gauge group $G$ and area $A$, it was shown in 
Ref.~\cite{FMS11} that $\tilde F_N(L)$ is related to $YM_2$ with the gauge 
group $G = {\rm Sp}(2N)$ via the relation
\begin{eqnarray}\label{relation_A_YM}
\tilde F_{N}(L) \propto {\cal Z}\left(A = \frac{2 \pi^2}{L^2}\, N, {\rm Sp}(2N) \right) \;.
\end{eqnarray}
In Ref. \cite{DK93,CNS96}, it was shown that for large $N$, ${\cal 
Z}(A,{\rm Sp}(2N))$ exhibits a third order phase transition at the 
critical value $A = \pi^2$ separating a weak coupling regime for $A < 
\pi^2$ and a strong coupling regime for $A > \pi^2$. This is the so called 
Douglas-Kazakov phase transition \cite{DK93}, which is the counterpart in 
continuum space-time, of the Gross-Witten-Wadia 
transition~\cite{GW80,Wad80} discussed above which is also of third order. 
Using the correspondence $L^2 = 2 \pi^2N/A$, we then find that $\tilde 
F_N(L)$, considered as a function of $L$ with $N$ large but fixed, also 
exhibits a third order phase transition at the critical value $L_c(N) = 
\sqrt{2N}$. Furthermore, the weak coupling regime ($A < \pi^2$) 
corresponds to $L > \sqrt{2N}$ and thus describes the right tail of 
$\tilde F_N(L)$, while the strong coupling regime corresponds to $L < 
\sqrt{2N}$ and describes instead the left tail of $\tilde F_N(L)$ (see 
Fig. \ref{fig_corres}). This is thus qualitatively very similar to the 
stability diagram of model (\ref{linear_stability}) depicted in Fig. 
\ref{fig:stability}. The critical regime around $A = \pi^2$ is the so 
called ``double scaling'' limit in the matrix model and has width of order 
$N^{-2/3}$. It corresponds to the region of width ${\cal O}(N^{-1/6})$ 
around $L = \sqrt{2N}$ where $\tilde F_N(L)$, correctly shifted and 
scaled, is described by the Tracy-Widom distribution~${\cal F}_1(x)$ \cite{FMS11,Lie12} given 
in (\ref{TW_124}).

In Refs. \cite{FMS11,SMCF} the three regimes for $\tilde F_N(L)$ (the left 
tail, the central part and the right tail) were studied in detail, 
yielding the results

\begin{eqnarray}\label{main_results_F}
\hspace*{-2.5cm}{\cal P}(H_N=L) \approx
\begin{cases}
&\exp{\left[-N^2 \phi_-^A\left(L/\sqrt{2N} \right) \right]} \;, \; L < \sqrt{2N} \; \& \; |L- \sqrt{2N}| \sim {\cal O}(\sqrt{N}) \\
& \\
&2^{11/6} N^{1/6}{\cal F}'_1\left[2^{11/6} N^{1/6} (L-\sqrt{2N}) \right] \;, \; |L-\sqrt{2N}| \sim {\cal O}(N^{-1/6}) \\
& \\
&\exp{\left[-N \phi_+^A\left(L/\sqrt{2N} \right) \right]} \;, \; L > \sqrt{2N} \; \& \; |L- \sqrt{2N}| \sim {\cal O}(\sqrt{N}) \;,\\
\end{cases}
\end{eqnarray}
where ${\cal F}_1$ is the TW distribution for GOE (\ref{TW_124}). The rate 
functions $\phi_\pm^A(x)$ can be computed exactly \cite{SMCF}: of 
particular interest are their asymptotic behaviors when $L \to \sqrt{2 N}$ 
from below (left tail) and from above (right tail), which are given by
\begin{eqnarray}\label{asympt_rate}
\phi_-^A(x) &\sim& \frac{16}{3}(1-x)^3 \;, \; x \to 1^- \;, \\
\phi_+^A(x) &\sim& \frac{2^{9/2}}{3}(x-1)^{3/2}\;, \; x \to 1^+ \;. \nonumber
\end{eqnarray}

The different behavior of the CDF $\tilde F_N(L)$ of the maximal height in 
the Brownian motion problem (\ref{main_results_F}) is thus formally very 
similar to the behavior of the CDF of $\lmax$ for Gaussian random matrices 
(\ref{regimes_gaussian}). In addition, the behavior of $\tilde F_N(L)$ for 
$L < \sqrt{2N}$ and $L > \sqrt{2N}$ leads also here, in the limit $N \to 
\infty$, to a phase transition at the critical point $L = \sqrt{2N}$ in 
the following sense. Indeed if one scales $L$ by $\sqrt{2N}$, keeping the 
ratio $x = L/\sqrt{2N}$ fixed, and take the limit $N \to \infty$ one 
obtains

\begin{eqnarray}\label{third_order_pt}
\lim_{N \to \infty} -\frac{1}{N^2} \ln \tilde F_N\left(x = \frac{L}{\sqrt{2N}} \right) = 
\begin{cases}
&\phi_-^A(x)\; , \quad x < 1 \\
&0 \quad\quad\quad x > 1 \;.
\end{cases}
\end{eqnarray}

If one interprets $\tilde F_N(L)$ in (\ref{Fp}) as the partition 
function of a discrete Coulomb gas, its logarithm can be interpreted as 
its free energy. Since $\phi_-^A(x) \sim (1-x)^3$ when $x$ approaches $1$ 
from below, then the third derivative of the free energy at the critical 
point $x = 1$ is discontinuous, which can also be interpreted as a third 
order phase transition, similar (albeit with different details) to the one 
found for 
the largest eigenvalue of Gaussian random matrices (\ref{asympt_phim}, \ref{transition_lmax}). Notice also that, thanks to the asymptotic behaviors 
of the rate functions (\ref{asympt_rate}), one can show that the matching 
between the different regimes is similar to the one found for $\lmax$ and 
discussed above in section \ref{section_matching}.

\subsection{Conductance and shot noise in mesoscopic cavities}

The second example concerns the large deviation formulas for linear 
statistics of the $N$ transmission eigenvalues $T_i$ of a chaotic cavity, 
in the framework of RMT. We consider here the statistics of quantum 
transport of electrons through a mesoscopic cavity, such as a quantum dot. 
This chaotic cavity is connected to two identical leads, each supporting 
$N$ channels. An electron, injected in the cavity through one lead, gets 
scattered in the cavity and leaves it by either of the two leads. In the 
Landauer-B\"uttiker approach \cite{Bee97,Lan57,But00}, the transport of 
electrons through such an open quantum system is encoded by the $2N\times 
2N$ unitary scattering matrix
\begin{eqnarray}
S = 
\begin{pmatrix}
r & t' \\
t & r' 
\end{pmatrix} \;,
\end{eqnarray}  
where the transmission $(t,t')$ and reflection $(r,r')$ blocks are $N 
\times N$ matrices encoding the transmission and reflection coefficients 
among different channels. Several physically relevant transport 
observables, such as the conductance ${G}$, or the shot noise power ${P}$ 
can be expressed in terms of the transmission eigenvalues $T_i$'s of the 
$N\times N$ matrix $T = tt^\dagger$. One has, for instance, for ${G}$ 
\cite{Lan57} and ${P}$ \cite{But00}
\begin{eqnarray}
{G} = \Tr(tt^\dagger) = \sum_{i=1}^N T_i \;, \; {P} = \Tr(tt^\dagger(1-tt^\dagger)) = \sum_{i=1}^N T_i(1-T_i) \;,
\end{eqnarray}
where $0 \leq T_i \leq 1$ denotes the probability that an electron gets 
transmitted through the channel $i$. Over the past two decades, RMT has 
been successfully used \cite{Bee97} to model the transport through such a 
cavity. Within this approach, one assigns a uniform probability density to 
all scattering matrices $S$ belonging to the unitary group: the matrix $S$ 
is thus drawn from one of Dyson's circular ensembles \cite{Meh91,For10}. 
It is then possible, though non trivial, to derive the joint PDF of the 
transmission eigenvalues $T_i$ which reads \cite{Bee97}:
\begin{eqnarray}
\tilde P_{\rm joint}(T_1, \cdots, T_N) = \tilde B_N(\beta) \Delta^\beta_N(T_1, \cdots, T_N) \prod_{i=1}^N T_i^{\frac{\beta}{2}-1}\;, \; 0\leq T_i \leq 1 \; \forall i \;,
\label{jpdf_cond}
\end{eqnarray}
where $\tilde B_N(\beta)$ is a normalization constant and the Dyson index 
characterizes the different symmetry classes ($\beta = 1,2$ according to 
the presence or absence of time reversal symmetry and $\beta = 4$ in case 
of spin-flip symmetry).

While formal expressions for the distributions of the conductance 
${\cal P}_N(G)$ and of the shot 
noise power ${\cal P}_N(P)$, for arbitrary finite $N$ and $\beta$,
were obtained in Refs.~\cite{Sommers1,Sommers2} and an exact recursion
relation for the cumulants of the conductance distribution was 
obtained in Ref.~\cite{OK2008}, it is not easy to 
obtain the large $N$ asymptotics of these results: indeed,  
the asymptotic tails of the conductance distribution derived
in Ref.~\cite{OK2008} by extrapolation of these finite $N$ cumulants turned 
out to be incorrect (as was demonstrated in \cite{VMB08,VMB10}). It turns out 
that an
easier method to derive directly the large $N$ results is
via using the Coulomb gas method~\cite{VMB08,VMB10}, where $T_i$'s (distributed 
via
the joint PDF (\ref{jpdf_cond})) can be interpreted as the
position of the $i$-th charge confined in a finite box $T_i\in [0,1]$, 
repelling
each other via the Vandermonde term and each subjected to an
external potential. The probability density of any linear statistic of $T_i$'s 
can then
be analyzed by performing a saddle point analysis of the underlying
Coulomb gas in the large $N$ limit~\cite{VMB08,VMB10}. Skipping details,
it was found that for both distributions ${\cal P}_N(G)$ and ${\cal P}_N(P)$,
there is a central Gaussian region flanked on both sides by non-Gaussian 
tails. For the conductance, ${\cal P}_N(G)$ behaves as
\begin{eqnarray}\label{behavior_conductance}
\hspace*{-2cm}&&{\cal P}_N(G) \approx \exp{\left[-\frac{\beta}{2} N^2 \Psi_G\left(\frac{G}{N} \right) \right]} \;, \; \Psi_G(x) =
\begin{cases}
\frac{1}{2} - \ln{(4x)} \;, \; &0 \leq x \leq \frac{1}{4} \; \\
8\left(x-\frac{1}{2}\right)^2 \;, &\frac{1}{4} \leq x \leq \frac{3}{4} \\
\frac{1}{2} - \ln{[4(1-x)]} \;, \; &\frac{3}{4} \leq x \leq 1 
\end{cases}\;,
\end{eqnarray}
and a similar (with slightly different details) behavior was found for 
${\cal P}_N(P)$, with an associated rate function 
$\Psi_P(x)$. It turns out that at the two singular points
$G/N=1/4$ and $G/N=3/4$, the underlying saddle point charge density
$\rho^*(x)$ changes form. Exactly at the two critical 
points, one edge of the Coulomb gas with density vanishing at the edge
as a square-root hits the hard physical boundary of the bounding box at $0$ and 
$1$ respectively~\cite{VMB08,VMB10}.
Consequently, the system undergoes a third order phase transition
at $G/N=1/4$ and also at $G/N=3/4$.
Indeed, one can check from 
(\ref{behavior_conductance}) that the third derivative of the rate 
function $\Psi_G(x)$ is discontinuous at these critical points. Such a third 
order phase transition also occurs for the distribution of the shot noise 
power \cite{VMB08} and is expected to occur for any generic linear 
statistics of the eigenvalues $T_i$'s \cite{VMB10}. 
These non-analyticities of the rate functions thus appear as a 
direct consequence of phase transitions in the associated Coulomb gas 
problems. The existence of these three different regimes 
(\ref{behavior_conductance}) were confirmed by numerical simulations 
\cite{VMB08,VMB10}. Finally, we note that 
similar large deviations regimes [albeit with an additional fourth regime 
compared to (\ref{behavior_conductance})] and associated third order phase 
transitions were found for the Andreev conductance of a 
metal-superconductor interface in Ref.~\cite{DMTV11}.

\subsection{Bipartite entanglement of a random pure state}

Another example of a third order phase transition is provided by the 
distribution of the bipartite 
entanglement of a random pure state. We consider a bipartite quantum 
system which is described by the tensorial product of two smaller Hilbert 
spaces ${\cal H}_A \otimes {\cal H}_B$. We denote by $N$ and $M$ the 
dimensions of ${\cal H}_A$ and ${\cal H}_B$ and introduce $c = N/M$, with 
$0< c \leq 1$. We are mainly interested in the case where both $M$ and $N$ 
are large. The limit $c=1$ corresponds to $M=N$ while $c \to 0$ 
corresponds to the case where $B$ is the environment (e.g. a thermostat) 
and $A$ the system of interest. Here we focus on the case $c=1$. We 
suppose that the full system $A \otimes B$ is described by a random pure 
state $|\psi \rangle$ (such that $\langle \psi | \psi \rangle = 1$) and we 
denote by $\rho = |\psi \rangle \langle \psi |$ the associated density 
matrix, satisfying $\Tr [\rho] = 1$. We are interested in the entanglement 
entropy and hence we consider the reduced density matrices $\rho_A = \rm 
\Tr_B[\rho]$ and $\rho_B = \rm \Tr_A[\rho]$. These two matrices $\rho_A$ 
and $\rho_B$ share the same non-negative eigenvalues $\lambda_1, \cdots, 
\lambda_N$ with the normalization constraint $\sum_{i=1}^N \lambda_i = 1$. 
If we denote by $|\lambda_i^A\rangle$ and $|\lambda_i^B\rangle$ the 
corresponding eigenvectors of $\rho_A$ and $\rho_B$, an arbitrary pure 
state $|\psi \rangle$ can be written in the so called Schmidt basis:
\begin{eqnarray}\label{schmidt_decomposition}
|\psi \rangle = \sum_{i=1}^N \sqrt{\lambda_i} |\lambda_i^A\rangle \otimes |\lambda_i^B\rangle \;.
\end{eqnarray}

As an example, let us consider two limiting cases: (i) if $\lambda_j = 1$ 
and the remaining $N-1$ eigenvalues are identically zero, then $|\psi 
\rangle = \lambda_j |\lambda_j^A\rangle \otimes |\lambda_j^B\rangle$. 
Hence the state factorizes and the system is completely unentangled. (ii) 
If instead all the eigenvalues are equal, $\lambda_i = 1/N$ for all $i$, 
all the states are equally represented in (\ref{schmidt_decomposition}) 
and the state is maximally entangled. A standard measure of entanglement 
is provided by the von Neumann entropy, $S_{\rm VN} = - \sum_{i=1}^N 
\lambda_i \ln \lambda_i$ [it takes its minimum value $0$ in case (i) and 
its maximal value $\ln N$ in case (ii)]. Another useful measure of 
entanglement is provided by the Renyi's entropies
\begin{eqnarray}\label{def_reny}
S_q = \frac{1}{q-1} \ln \Sigma_q \;, \; \Sigma_q=\sum_{i=1}^N \lambda_i^q \;,
\end{eqnarray}
and we restrict here our analysis to the case $q \geq 1$. $S_q$ is also minimal in case (i) where $S_q = 0$ and maximal for case (ii) where $S_q = \ln N$.  Note that $S_q \to S_{\rm VN}$ when $q \to 1$ while $S_q \to -\ln \lmax$ when $q \to \infty$ (where $\lmax = \max_{1\leq i \leq N} \lambda_i$). If the random pure state $|\psi\rangle$ is uniformly distributed (i.e. according to the uniform Haar measure) the eigenvalues $\lambda_i$'s are also random variables with the joint PDF given by \cite{Pag93}
\begin{eqnarray}\label{joint_entropy}
{\cal P}^W_{\rm joint}(\lambda_1, \cdots, \lambda_N) = {\cal B}^W_N(\beta) \prod_{i=1}^N \lambda_i^{\frac{\beta}{2}-1} \Delta^\beta_N(\lambda_1, \cdots, \lambda_N) \delta\left(\sum_{i=1}^N \lambda_i - 1\right) \;,
\end{eqnarray}
where $\Delta_N(\lambda_1, \cdots, \lambda_N)$ is the Vandermonde 
determinant (\ref{vdm}) and ${\cal B}^W_N(\beta)$ a normalization 
constant. Here $\beta = 2$ and the $\delta$-function enforces the 
constraint that ${\Tr} [\rho_A] =1$. Note that, apart from this 
constraint, this joint PDF (\ref{joint_entropy}) is identical to the 
eigenvalue distribution of random Gaussian Wishart matrices 
(\ref{joint_wishart}).

In Refs.~\cite{NMV10,NMV11}, the PDF of $\Sigma_q$ [which yields the PDF 
of $S_q$ itself from (\ref{def_reny})] was computed using Coulomb gas 
techniques similar to the ones explained above in 
section~\ref{section:CG} for all $q$. 
For the special case $q=2$, the Laplace transform of the distribution
of purity $\sum_{i=1}^N \lambda_i^2$ was studied by similar methods
in Refs.~\cite{Facchi08,Facchi10}.
The main results obtained in Refs. 
\cite{NMV10,NMV11} can be summarized as follows. First, due to the global 
constraint, $\sum_{i=1}^N \lambda_i = 1$, one deduces that the typical 
scale of $\lambda_i$'s is ${\cal O}(1/N)$ and hence $\Sigma_q \sim {\cal 
O}(N^{1-q})$. Note also that for $q\geq 1$, one has necessarily $N^{1-q} 
\leq \Sigma_q \leq 1$. It was further shown in Refs. \cite{NMV10,NMV11} 
that the PDF ${\cal P}(\Sigma_q = N^{1-q}s)$ displays three different 
regimes: the distribution has indeed a Gaussian peak [centered on the mean 
value $\bar s(q)$] flanked on both sides by two non-Gaussian tails. There 
are thus three different regimes separated by two critical values $s_1(q)$ 
and $s_2(q)$ such that
\begin{eqnarray}\label{results_entanglement}
{\cal P}(\Sigma_q = N^{1-q}s) \approx 
\begin{cases}
&\exp{\left[-\beta N^2 \Phi_I(s)\right]} \;, \;  1 \leq s \leq s_1(q) \\
\\
&\exp{\left[-\beta N^2 \Phi_{II}(s)\right]} \;, \; s_1(q) \leq s \leq s_2(q) \\
\\
&\exp{\left[-\beta N^{1+\frac{1}{q}} \Phi_{III}(s)\right]} \;, \; s > s_2(q) 
\;.
\end{cases}
\end{eqnarray}
Interestingly, at the first critical point 
$s_1(q)$, the PDF exhibits also a third-order phase transition (i.e. the 
third derivative of the large deviation function is discontinuous), as 
found above. On the other hand, at the second critical point $s_2(q)$, a 
Bose-Einstein type condensation occurs and the distribution changes shape 
abruptly: in this case the first derivative is discontinuous in the limit 
$N \to \infty$. Here also, these changes of behavior 
(\ref{results_entanglement}) are a direct consequence of two phase 
transitions in the associated Coulomb gas problem, and more precisely in 
the shape of the optimal charge density~\cite{NMV10,NMV11}.
 
\section{Basic mechanism of the third order transition and its generalizations}

\begin{figure}\vspace*{3cm}
\begin{center}
\includegraphics[width = 0.7\linewidth]{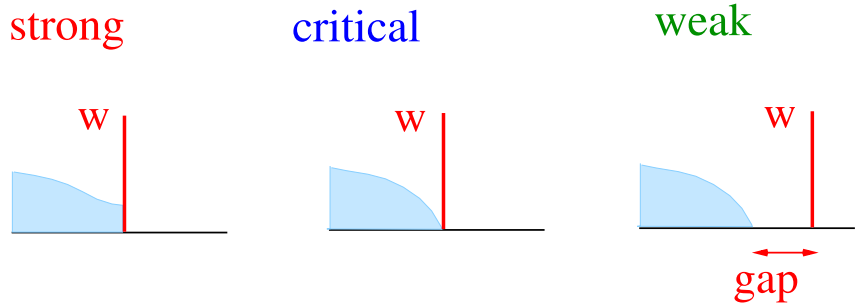}
\caption{A third order phase transition between a weak coupling
and a strong coupling phase occurs when the 
gap, between
the soft edge of a Coulomb droplet with density vanishing as a square-root
at the edge and a hard wall at $w$, vanishes as one tunes a control parameter
through its critical value.}
\label{fig:gap}
\end{center}
\end{figure}

All the problems discussed so far in this article share one common feature: 
there is a third order phase transition as a control parameter $\alpha$
crosses a critical value $\alpha_c$. In the case of May's model of dynamical
systems, $\alpha$ denotes the strength of the pairwise interaction between
species, whereas in the $2$-d lattice QCD, $\alpha=g$ is the coupling strength
of the gauge fields. The basic mechanism behind this third order phase 
transition can be summarized as follows. In all these problems, there is
an underlying one dimensional Coulomb gas, with charge density
supported over a {\em single} interval and in presence 
of a hard wall located
at $w$. The equilibrium charge 
density of the Coulomb gas
$\rho_{\alpha}(x)$ depends
on $\alpha$.  In the {\it weak coupling} phase 
($\alpha<\alpha_c$), $\rho_\alpha(x)$ has typically a soft edge at 
$x=b<w$ where the density vanishes as a square root, 
$\rho_{\alpha}(x)\sim (b-x)^{1/2}$. 
Thus, in this case, there is a nonzero gap between the soft edge
of the Coulomb gas and the hard wall at $w$, and the charges do not feel
the presence of the wall. As $\alpha$ increases and approaches the critical
value $\alpha_c$, the soft edge approaches the hard wall leading to
a vanishing gap. For $\alpha>\alpha_c$, the edge of the charge density
gets {\it pushed} by the wall and the systems adjusts itself to
a new deformed equilibrium charge density with a nonzero
density at the wall location--this is the so called
{\it strong coupling} phase ($\alpha>\alpha_c$) (see Fig.~\ref{fig:gap}).

In all these cases where the charge density vanishes as a square-root 
at the soft edge of the single support in the weak coupling phase, one 
obtains
a third order phase transition from the weak to strong coupling phase when the 
tuning parameter crosses the
critical value $\alpha=\alpha_c$ from below. For finite but large $N$, if one
zooms in the critical region, one finds a smooth crossover from the
weak coupling to the strong coupling phase and the crossover function
has the Tracy-Widom scaling form (expressed in terms of a solution of a Painlev\'e-II differential
equation). In fact, in the context of $2$-d QCD, this was already noticed
by Periwal and Shevitz~\cite{PS90} before the work of Tracy and 
Widom~\cite{TW94}, but there was no probabilistic 
interpretation
of this crossover function. From the work of Tracy and Widom~\cite{TW94}, it 
follows
that this crossover function can be interpreted as the centered and scaled 
limiting distribution of the largest eigenvalue of a Gaussian random matrix.

What happens if the equilibrium charge density vanishes at the soft edge
(in the weak coupling phase) not as a square-root, but say as $\tilde \rho(x)\sim
(b-x)^{\gamma}$ with an exponent $\gamma>0$~? How does the order of the 
transition depend on $\gamma$~?
Indeed, this question arises in the context of higher order critical matrix 
models~\cite{BIPZ}. 
Consider
for instance a random $(N\times N)$ matrix ${X} $ whose entries are drawn 
from the joint 
distribution ${\rm Prob.}[X]\propto \exp\left[- N\, {\rm Tr}\left(V( 
X)\right)\right]$ where $V(X)$ is a polynomial potential. By choosing
the potential $V(X)$ appropriately, one can engineer equilibrium charge 
densities that are different from the Wigner semi-circular form~\cite{BIPZ}.
For instance, if one chooses $V(X)= X^4/20-4X^3/15+X^2/5+8X/5$ and with
$\beta=2$, the
saddle point charge density can be computed exactly~\cite{CIK09,CO11}
\begin{eqnarray}
\tilde \rho(x)= \frac{1}{10\pi} (x+2)^{1/2} (2-x)^{5/2}, \quad x\in [-2,2] \;,
\label{quartic.1}
\end{eqnarray}
which thus vanishes with an exponent $\gamma=5/2$ at the upper soft edge
$b=2$. For such matrix models with a higher order critical point, the
distribution of the typical fluctuations of the largest eigenvalue around its 
mean $b$ are described by higher-order analogues of the Tracy-Widom 
distribution~\cite{CIK09,CO11}. One also expects that there
would be the analogues of the large deviation functions of $\lmax$, just as 
in the simple quadratic case $V(X)=X^2/2$ discussed in Section 2
and these large deviation rate functions (both left and right) have recently 
been computed
using generalized orthogonal polynomial 
techniques~\cite{Akemann_Atkin,Atkin_Zohren}.
Clearly there would be a phase transition in the CDF of $\lmax$ at the critical
value $\lmax=b$ for these critical matrix models as well. What is the order of 
this phase transition? This order can be
easily estimated by the following simple scaling argument.

Let, in general, $\tilde \rho(x) \sim (b-x)^{\gamma}$ at the upper soft edge $x=b$. 
One can
easily estimate the scale of typical fluctuation $\delta \lambda_{\max}$ of 
$\lmax$ around its mean 
$b$. Using the standard EVS criterion [see Eq. (\ref{EVS.1})], i.e.,
setting $\int_{b-\delta \lambda_{\max}}^{b} \tilde \rho(x) dx\sim 1/N$, one
gets
\begin{eqnarray}
\delta {\lmax} = b - \lmax \sim {\cal O}(N^{-1/(1+\gamma)})  \;.
\label{typ_multi.1}
\end{eqnarray}  
For $\gamma=1/2$, one recovers $\delta {\lmax}\sim {\cal O}(N^{-2/3})$. Hence, 
one would 
expect that on this scale, the CDF of $\lmax$ will have the scaling form
\begin{eqnarray}
{\rm Prob.}[\lmax\le w]\sim {\cal F}\left(N^{1/(1+\gamma)}(w-b)\right) \;,
\label{high_tw.1}
\end{eqnarray}
where the scaling function ${\cal F}(x)$ is the $\gamma$-analogue of the 
Tracy-Widom
function. Now, in general, we would expect that far in the left tail, this
function should decay asymptotically as,
\begin{eqnarray}
{\cal F}(x)\sim \exp[- a_0\, |x|^{\delta}] \;, \; {\rm for \;} x \to -\infty \;,
\label{left_tw.1}
\end{eqnarray}  
where $a_0$ is a constant. Clearly, for $\gamma=1/2$ case (i.e., when 
${\cal F}(x)$
is the standard Tracy-Widom), one has $\delta=3$ [see Eq. (\ref{asympt_TW})].

On the other hand, it follows from the general
Coulomb gas
argument in Section 3 that 
atypical fluctuations of $\lmax$ of $\sim {\cal O}(1)$ to the 
left of $b$, i.e., when $w<b$, are described by a large deviation
form
\begin{eqnarray}
{\rm Prob.}[\lmax\le w]\sim \exp\left[-\beta\, N^2 \Phi_{-}(w)\right], \quad 
w<b \;,
\label{multi_left.1}
\end{eqnarray}
where $\Phi_{-}(w)$ is a rate function that should vanish as $w\to b$ from 
the left. Interpreting $\Phi_{-}(w)$ as the free energy as in the Gaussian 
case, we then expect $\Phi_{-}(w)\sim a_1\, (b-w)^{\sigma}$ as $w\to b$
where $a_1$ is a constant and the exponent
$\sigma$ then decides the order of the transition. To estimate $\sigma$,
we match this left large deviation results (when $w\to b$) with the extreme 
left tail of the central peak region as described in (\ref{left_tw.1}).
Substituting $\Phi_{-}(w)\sim a_1\, (b-w)^{\sigma}$ in (\ref{multi_left.1})
gives, for $w\to b$,
\begin{eqnarray} 
{\rm Prob.}[\lmax\le w]&\sim & \exp\left[- \beta\, N^2\, a_1\, 
(b-w)^{\sigma} \right] \nonumber \;,\\
&\sim & \exp\left[-\beta\, a_1\, \left[N^{2/\sigma}\, 
(b-w)\right]^{\sigma}\right] \;.
\label{multi_left.2}
\end{eqnarray}
In contrast, for $(b-w) \gg N^{-1/(1+\gamma)}$, we get, by using the
left tail asymptotics (\ref{left_tw.1})
of the central peak behavior in (\ref{high_tw.1}),
\begin{eqnarray}
{\rm Prob.}[\lmax\le w]&\sim & \exp\left[-a_0\, 
{\left\{ N^{1/(1+\gamma)}\,(b-w)\right\} }^{\delta}\right]\;.
\label{left_tw.2}
\end{eqnarray}
Assuming that the two behaviors merge smoothly, we find by comparing 
(\ref{multi_left.2}) and (\ref{left_tw.2})
\begin{eqnarray}
\delta=\sigma \;\; {\rm and}\;\; \frac{\delta}{1+\gamma}=2 \;,
\label{match.2}
\end{eqnarray} 
which then relates the order of the transition $\sigma$ to the exponent 
$\gamma$ 
characterizing
the vanishing of the charge density at the soft edge, via the simple scaling 
relation
\begin{eqnarray}
\sigma= 2\,(1+\gamma)\, .
\label{scaling_relation}
\end{eqnarray}
For example, for $\gamma=1/2$, one recovers the third order transition
$\sigma=3$. As an example, the case (\ref{quartic.1}) with $\gamma=5/2$,
will then have a seventh order ($\sigma=7$) phase transition.

\section{Conclusion}

To summarize, we have provided a rather complete understanding of the
behavior of the probability distribution of the top eigenvalue
$\lmax$ of an $N\times N$ Gaussian random matrix for large $N$.
While the typical small fluctuations of $\lmax$ of order $\sim {\cal O}(N^{-2/3})$ 
around its mean $\lmax=\sqrt{2}$ are described by Tracy-Widom distributions,
atypically large fluctuations of $\sim {\cal O}(1)$ are described by
two different large deviation functions respectively on the left and on
the right of the mean. These two tails correspond to very different physics
in terms of the underlying Coulomb gas describing the eigenvalues:
the left large deviation corresponds to a {\it pushed} Coulomb gas,
while the right large deviation corresponds to a {\it pulled} Coulomb gas. 
We have shown that the left large deviation
function can be interpreted as a thermodynamic free energy and that 
it undergoes a third order phase transition at $\lmax=\sqrt{2}$, i.e,
its third derivative is discontinuous at $\lmax=\sqrt{2}$. This
result provides a thermodynamic meaning to the stable-unstable
phase transition in May's model of dynamical systems. Furthermore,
it shows that this phase transition occurs in a broad class of systems ranging
from dynamical systems all the way to two dimensional gauge theory. 
The {\it pushed} phase (i.e., the {\it unstable} phase in May's model)
corresponds to the {\it strong coupling} phase of the gauge theory, while
the {\it pulled} phase (i.e., the {\it stable} phase in May's model)
corresponds to the non-perturbative {\it weak coupling} phase of 
the gauge theory. In addition, several other physical systems
with a similar third order phase transition have been discussed.
The
basic mechanism behind this transition is also identified: it happens
when the gap between the soft edge characterizing the equilibrium 
charge density of an underlying Coulomb gas with a single support and a 
hard wall vanishes.
The generalizations to higher order phase transitions have also been discussed.

The main interesting point to note is that the large deviation function
associated with the probability distribution of an observable $\hat O$ of an 
$(N\times N)$  random matrix, such 
as the largest 
eigenvalue 
$\hat O=\lmax$ or other
linear statistics of the form $\hat O =\sum_{i=1}^N f(\lambda_i)$ (where
$f(x)$ is an arbitrary function), may exhibit singularities in the large $N$
limit and these singularities typically signal a phase transition
in the underlying Coulomb gas. Hence, these large deviation functions
are indeed the analogues of the thermodynamic free energy 
in a standard statistical mechanical system. Here we have discussed
several cases where these singularities are of the power-law variety,
i.e., vanishes as some power near the critical point. However, other
cases with essential singularities~\cite{NM_interface} and logarithmic 
singularities~\cite{index.1,index.2,index.3} have also been 
identified. In addition, in several examples a first order
phase transition (akin to Bose-Einstein condensation) has been shown to take 
place when a single eigenvalue 
splits off the sea of eigenvalues~\cite{NMV11,TM13}.
Finally, we note that recently a third order phase transition has been 
found
in two dimensions~\cite{ATW2013}--in the large deviation function 
associated with
the index distribution in real and complex Ginibre random matrices. The 
mechanism for this third order transition in two dimensions (where the
charge density in the complex plane has two supports)
seems to be different from the one dimensional cases with
a single support discussed in this article.

In conclusion, the large deviation functions associated with
the probability distribution of an observable in RMT carry crucial
informations concerning phase transitions in the system in the form of 
singularities and hence are useful and important objects to study.

\ack

This review is dedicated to the memory of Oriol Bohigas from whom
we learnt many aspects of random matrix theory.
We would also like to thank R. Allez, G. Akemann, J.~Baik, G. Ben Arous, O. 
Bohigas, G. Borot, 
J.-P. 
Bouchaud, A. 
Comtet, K.~Damle, D. S. Dean, D. Dhar, B. Eynard, P. J. Forrester, 
A.~Lakshminarayan, C. Nadal, S. Nechaev, J. Rambeau, A.~Scardicchio, C. 
Texier, S. Tomsovic, V. 
Tripathi, M. Vergassola, D. Villamaina, P. Vivo, G. Wainrib, O. Zeitouni 
for useful discussions and 
collaborations. We acknowledge support by ANR grant 
2011-BS04-013-01 WALKMAT and in part by the Indo-French Centre for the 
Promotion of Advanced Research under Project 4604-3.

\clearpage


\end{document}